\documentclass[a4paper,longbibliography,english,reprint,aps,superscriptaddress,nofootinbib]{revtex4-2}

\usepackage{babel,calc,amsmath,amsthm,amssymb,graphicx,subfigure,xcolor}
\usepackage[T1]{fontenc}
\usepackage{mathdots}
\usepackage{geometry}
\usepackage[unicode=true]{hyperref}
\usepackage{booktabs}
\usepackage{mathtools}
\usepackage{multirow}
\usepackage{lipsum}
\usepackage[normalem]{ulem}

\usepackage{multibib}

\definecolor{linkblue}{rgb}{0.2314, 0.4118, 0.6196}
\hypersetup{
     colorlinks=true,       		
     linkcolor=linkblue,          	
     citecolor=linkblue,             
     urlcolor=linkblue,          
 }

\newcommand{\ket}[1]{\left\vert#1\right\rangle}
\newcommand{\bra}[1]{\left\langle#1\right\vert}
\newcommand{\braket}[1]{\left\langle#1\right\rangle}

\newcommand{\tr}{\rm tr}

\newtheorem*{corollary*}{Corollary}

\newtheorem*{proposition*}{Proposition}
\newtheorem{theorem}{Theorem}
\newtheorem*{theorem*}{Theorem}

\theoremstyle{definition}

\newtheorem*{definition*}{Definition}

\theoremstyle{remark}

\newtheorem*{lemma*}{Lemma}

\newtheorem*{remark*}{Remark}

\newtheorem*{example*}{Example}

\usepackage{tipa}
\newcommand{\revc}{\ensuremath{\textbf{\textopeno}}}

\newif\ifdebug

\ifdebug
\definecolor{zhliu}{rgb}{0.60, 0.12, 0.24}
\newcommand{\add}[1]{{\color{zhliu}{#1}}}
\newcommand{\note}[1]{{\color{orange}{#1}}}
\newcommand\delete{\bgroup\markoverwith{\textcolor{zhliu}{\rule[0.5ex]{2pt}{0.8pt}}}\ULon}

\else
\newcommand{\add}[1]{#1}
\newcommand{\note}[1]{\ignorespaces}
\newcommand{\delete}[1]{\ignorespaces}

\fi

\DeclareSymbolFont{extraup}{U}{zavm}{m}{n}
\DeclareMathSymbol{\varspade}{\mathalpha}{extraup}{85}

\begin{document}
\renewcommand{\figurename}{Fig.}

\newcommand{\extfig}{Extended Data Figure}
\newcommand{\methodsname}{Methods}
\newcommand{\smname}{Supplementary Material}

\title{\vspace{-12pt}\mbox{\hspace{-10pt}Exploring the boundary of quantum correlations with a time-domain optical processor}}

\affiliation{CAS Key Laboratory of Quantum Information, University of Science and Technology of China, Hefei 230026, China}
\affiliation{School of Mathematical Sciences, University of Science and Technology of China, Hefei 230026, China}
\affiliation{CAS Centre for Excellence in Quantum Information and Quantum Physics, University of Science and Technology of China, Hefei 230026, China}
\affiliation{Beijing QBoson Quantum Technology Co., Ltd., Beijing 100015, China}
\affiliation{School of Physics and Optoelectronics Engineering, Anhui University, 230601 Hefei, People’s Republic of China}
\affiliation{Naturwissenschaftlich-Technische Fakult\"{a}t, Universit\"{a}t Siegen, Walter-Flex-Stra\ss e 3, 57068 Siegen, Germany}
\affiliation{Quantum Science Center of Guangdong-Hong Kong-Macao Greater Bay Area, Shenzhen 518045, China}
\affiliation{Theoretical Physics Division, Chern Institute of Mathematics, Nankai University, Tianjin 300071, China}
\affiliation{Anhui Province Key Laboratory of Quantum Network, University of Science and Technology of China, Hefei, Anhui 230026, China}
\affiliation{Hefei National Laboratory, University of Science and Technology of China, Hefei 230088, China}

\author{Zheng-Hao~Liu
$^{\circ,\#}$
}
\affiliation{CAS Key Laboratory of Quantum Information, University of Science and Technology of China, Hefei 230026, China}
\affiliation{CAS Centre for Excellence in Quantum Information and Quantum Physics, University of Science and Technology of China, Hefei 230026, China}
\affiliation{Beijing QBoson Quantum Technology Co., Ltd., Beijing 100015, China}

\author{Yu~Meng
$^{\lozenge,\#}$
}
\affiliation{CAS Key Laboratory of Quantum Information, University of Science and Technology of China, Hefei 230026, China}
\affiliation{CAS Centre for Excellence in Quantum Information and Quantum Physics, University of Science and Technology of China, Hefei 230026, China}

\author{Yu-Ze~Wu
$^{\#}$
}
\affiliation{School of Mathematical Sciences, University of Science and Technology of China, Hefei 230026, China}

\author{Ze-Yan~Hao}
\affiliation{CAS Key Laboratory of Quantum Information, University of Science and Technology of China, Hefei 230026, China}
\affiliation{CAS Centre for Excellence in Quantum Information and Quantum Physics, University of Science and Technology of China, Hefei 230026, China}

\author{Zhen-Peng~Xu}
\email{zhen-peng.xu@ahu.edu.cn}
\affiliation{School of Physics and Optoelectronics Engineering, Anhui University, 230601 Hefei, People’s Republic of China}
\affiliation{Naturwissenschaftlich-Technische Fakult\"{a}t, Universit\"{a}t Siegen, Walter-Flex-Stra\ss e 3, 57068 Siegen, Germany}

\author{Cheng-Jun~Ai}
\author{Hai~Wei}
\affiliation{Beijing QBoson Quantum Technology Co., Ltd., Beijing 100015, China}

\author{Kai~Wen}
\email{wenk@boseq.com}
\affiliation{Beijing QBoson Quantum Technology Co., Ltd., Beijing 100015, China}
\affiliation{Quantum Science Center of Guangdong-Hong Kong-Macao Greater Bay Area, Shenzhen 518045, China}

\author{Jing-Ling~Chen}
\email{chenjl@nankai.edu.cn}
\affiliation{Theoretical Physics Division, Chern Institute of Mathematics, Nankai University, Tianjin 300071, China}

\author{Jie~Ma}
\email{jiema@ustc.edu.cn}
\affiliation{School of Mathematical Sciences, University of Science and Technology of China, Hefei 230026, China}
\affiliation{Hefei National Laboratory, University of Science and Technology of China, Hefei 230088, China}

\author{Jin-Shi~Xu}
\email{jsxu@ustc.edu.cn}

\author{Chuan-Feng~Li}
\email{cfli@ustc.edu.cn}

\author{Guang-Can~Guo}
\affiliation{CAS Key Laboratory of Quantum Information, University of Science and Technology of China, Hefei 230026, China}
\affiliation{CAS Centre for Excellence in Quantum Information and Quantum Physics, University of Science and Technology of China, Hefei 230026, China}
\affiliation{Anhui Province Key Laboratory of Quantum Network, University of Science and Technology of China, Hefei, Anhui 230026, China}
\affiliation{Hefei National Laboratory, University of Science and Technology of China, Hefei 230088, China}

\date{\today}

\begin{abstract}
    Contextuality is a hallmark feature of the quantum theory that captures its incompatibility with any noncontextual hidden-variable model. The Greenberger--Horne--Zeilinger (GHZ)-type paradoxes are proofs of contextuality that reveal this incompatibility with deterministic logical arguments. \add{However, the GHZ-type paradox whose events can be included in the fewest contexts and which brings the strongest nonclassicality remains elusive.}
    Here, we derive a GHZ-type paradox with a context-cover number of three and show this number saturates the lower bound posed by quantum theory. We demonstrate the paradox with a time-domain fiber optical platform and recover the quantum prediction in a 37-dimensional setup based on high-speed modulation, convolution, and homodyne detection of time-multiplexed pulsed coherent light. By proposing and studying a strong form of contextuality in high-dimensional Hilbert space, our results pave the way for the exploration of exotic quantum correlations with time-multiplexed optical systems.
\end{abstract}

\maketitle

\let\oldfootnote\thefootnote
\renewcommand{\thefootnote}{$\circ$}
\footnotetext{Present address: Center for Macroscopic Quantum States (bigQ), Department of Physics, Technical University of Denmark, Fysikvej, 2800 Kongens Lyngby, Denmark. \url{https://manekimeow.github.io/}
}
\renewcommand{\thefootnote}{$\lozenge$}
\footnotetext{Present address: Center for Hybrid Quantum Networks (Hy-Q), The Niels Bohr Institute, University of Copenhagen, DK-2100 Copenhagen \O, Denmark.}
\renewcommand{\thefootnote}{$\#$}
\footnotetext{These authors contributed equally to this work.}
\let\thefootnote\oldfootnote

\noindent\textsf{\textbf{\large Introduction}}

Measurement in quantum theory is the origin of many nonclassical effects\,\cite{EPR35,Wigner61,KS67}. A measurement of a physical property in the classical world reveals its preexisting value. Such an interpretation cannot hold true in the quantum regime, not only because noncommuting operators have no meaningful joint values\,\cite{Heisenberg27}, but also due to the impossibility of specifying the preexisting value, even for a set of commuting operators, without specifying the full set of observables jointly measured.
The latter effect, known as the Kochen--Specker contextuality\,\cite{note1}, captures some most defining aspects of quantum correlations. \add{It has a wide spectrum of applications like randomness expansion\,\cite{Abbott12,Um20}, dimension witnessing\,\cite{Guehne14,Ray21} and self-testing\,\cite{Bharti19,xmhu23}, and can be considered as the origin of nonclassicality behind Bell nonlocality\,\cite{Clifton93, Cabello21BKS}}.


\add{Intriguingly, the failure of noncontextual hidden-variable descriptions can manifest as a ``perfect'' form of contextuality, namely, the Greenberger--Horne--Zeilinger (GHZ)-type paradoxes\,\cite{GHZ89,Mermin90GHZ}, where the quantum system generates an outcome deterministically different from the noncontextual prediction. Such paradoxes formulated as inequality-free logical arguments serve as a conceptually clear and mathematically strong\,\cite{Abramsky11} alternative of noncontextuality inequalities for detecting contextuality. We argue that these perfect contextuality are also a resource for the acceleration and universality of quantum computing\,\cite{Anders09,Raussendorf13,Abramsky17,Howard14,Bravyi18,Bravyi20}. We refer our readers to the \smname\ for a more in-depth discussion about the role of contextuality in quantum computing.} 


The figure of merit of a GHZ-type paradox can be defined in several ways. It has been proved that the construction in~\cite{Cabello96} uses the fewest number of rays\,\cite{zpxu20peres} to show the paradox, and the number of contexts that must be utilized to prove the logical argument (the ``number of context'') is at least four\,\cite{Budroni21}. Here, we focus on an alternative point: what is the minimal number of contexts required to include, or ``cover'' all the events in a GHZ-type paradox? 
Hereafter, we call this quantity the ``number of context-cover'' to differentiate from the number of context. Later, the word choice will be made more reasonable. 
\add{The number of context-cover is a proper measure of the strength of a GHZ-type paradox for two reasons:}
firstly, when transformed into noncontextuality inequalities, a GHZ-type paradox with a lower number of context-cover yields a larger ratio of violation. This violation also sustains to the one of Kochen--Specker-type paradox~\cite{zpxu20peres}, a fundamental resource for various applications\,\cite{Howard14,miller2017universal,zhen2023device}. 
Secondly, to observe a GHZ-type paradox, the number of groups of fundamental event probabilities required is also equal to the number of context-cover. \add{Therefore, the GHZ-type paradox with the lowest number of context-cover is a strong resource of nonclassicality that is expected to have potential applications in quantum protocols}. However, after more than 30 years of GHZ's seminal work\,\cite{GHZ89}, the exact lower bound of this number is still undetermined. 

Here, we take a step towards answering the above question and observing this strong form of GHZ-type paradox. Specifically, we have constructed and demonstrated such a paradox with a number of context-cover of only three. Inspired by the graph-theoretic approach to quantum correlations\,\cite{CSW14}, we have developed a method to systematically construct GHZ-type paradoxes by searching for graphs whose graph-theoretic constants satisfy a set of criteria. Based on this method, we have explicitly constructed a three-context GHZ-type paradox that can be realized with a set of 37-dimensional measurements, and proved its optimality in the sense that the number of contexts to include all the events cannot be further reduced. This way, the proposed GHZ-type paradox delineates the boundary of correlations allowed by the quantum theory.

To experimentally study the three-context GHZ-type paradox, we have built a fiber-based photonic processor capable of reproducing the probability of all the high-dimensional measurements in the paradox with the time-bin degree of freedom of photons. Our experiment was based on high-speed electro-optical modulation, multiplication, and convolution on temporal modes of pulsed coherent light, with which we implemented a prepare-and-measure experiment through a correspondence between single-photon and coherent state interference\,\cite{Barnett22} and extracted all the statistics required in a test of contextuality\,\cite{Cabello16}. 
A major contribution from the experimental side is that we have made the setup scalable by detecting the complete amplitude and phase information of the coherent light via homodyne detection. This enabled us to substantially expand the applicable Hilbert space dimension and achieve the desired high-dimensionality. 

Our results epitomized the potential of the temporal-multiplexed optical system, which has found applications in coherent Ising machine\,\cite{McMahon16,Inagaki16,Honjo21}, continuous-variable measurement-based quantum computing\,\cite{Asavanant19,Larsen19,Larsen21,syren24}, and (Gaussian) boson sampling\,\cite{Xanadu22,Sempere22,yhe17}, in investigating exotic quantum correlations.
As any Kochen--Specker set utilizing $n$ context-cover implies the existence of a GHZ-type paradox with $n-1$ context-cover\,\cite{zpxu20peres}, our finding also paves the way to the search for a Kochen--Specker set whose events can be included in only four context-cover. We also envisage that our findings could benefit the search for exotic quantum correlations that are the building blocks for quantum computing or even constitute a step toward achieving stronger quantum advantage in shallow circuits\,\cite{Bravyi18,Bravyi20}.

\vspace{10pt}\noindent\textsf{\textbf{\large Results}}

\vspace{10pt}\noindent\textsf{\textbf{Graph-theoretic approach to GHZ-type paradox}}\hfill

The GHZ-type paradoxes are logical proofs and extreme manifestations of contextuality. They refer to scenarios where, given an assemblage of event probabilities, the predictions by quantum theory and any noncontextual model for another set of event probabilities contradict each other in a deterministic way. Formally, a GHZ-type paradox can be expressed using the conditional probabilities as follows:
\begin{align}
\begin{alignedat}{3}
    {\textstyle \sum_{k=1}^{m_1}}&\Pr(1|[1,k]) = 1, \\
    {\textstyle \sum_{k=1}^{m_2}}&\Pr(1|[2,k]) = 1, \\
    & \cdots \\
    {\textstyle \sum_{k=1}^{m_{n-1}}}&\Pr(1|[n-1,k]) = 1, \\[3pt]
    \midrule\midrule
    {\textstyle \sum_{k=1}^{m_n}}&\Pr(1|[n,k]) =\Big\{
    \begin{array}{ll}
        0, & \text{NCHV}, \\
        1, & \text{Q}.
    \end{array}
\end{alignedat}
\label{eq:ghz-general}
\end{align}
where $1|[j,k], j, k\in\mathbb{N^+}$ denotes an event where the outcome of the $k$-th projective measurement in the $j$-th context is 1, $m_j$ is the number of measurements in the $j$-th context, and $n$ is the total number of contexts used in the paradox. The qualifiers Q and NCHV indicate the probabilities shall be calculated via the quantum theory and a noncontextual hidden-variable model, respectively. We can also label all the events with a single index: $[j,k]=\sum_{i=1}^{j-1}m_i+k$. In layman's terms, a GHZ-type paradox indicates some events deemed impossible by noncontextual models are bound to happen according to the quantum theory. In the \smname, we provide an explicit illustration using the original GHZ paradox. 

To study the mathematical structure of contextuality, a common approach is the graph-theoretic approach to quantum correlations\,\cite{CSW14} which uses an exclusivity graph to capture the impossibility of some events taking place simultaneously. Concretely, the vertices $V(G)$ of the exclusivity graph $G$ represent the events of observing certain measurement outcomes, and its edges $E(G)$ connect pairs of exclusive events. Once an exclusivity graph is given, the sum of event probabilities in noncontextual and quantum theories will be bounded by graph constants, namely, the independence number and Lov\'asz number\,\cite{Cabello16,Lovasz79}:
\begin{align}
    \sum_{i\in V} \Pr(1|i) - \sum_{(i,j)\in E} \Pr(1,1|i,j) \overset{\rm NCHV}{\leqslant} \alpha(G) \overset{\rm Q}{\leqslant} \vartheta(G).
    \label{eq:CSW}
\end{align}
Here, $\Pr(1,1|i,j)$ is the probability of simultaneously observing events $i, j$; this term compensates for the deviation from exclusivity\,\cite{Cabello15}. Graphs with a large ratio $\vartheta/\alpha$ thus have the merit of producing more significant inconsistency between noncontextual and quantum theories. The advantage of the graph-theoretic approach is that many properties of the underlying quantum correlation can be revealed by the graph constants, and the concrete setting for the quantum states and measurements achieving these properties can then be found efficiently via semidefinite programming. 

To date, no direct method has, to the best of our knowledge, been established for studying the GHZ-type paradox via this approach. Here, we undertake this task and elucidate the link between a GHZ-type paradox and the clique number of its exclusivity graph, or equivalently, the \textit{chromatic number} of the graph with the rays in the GHZ-type paradox being its orthogonal representation\,\cite{Lovasz89}. Formally, our finding can be formulated as the following theorem.
\begin{theorem}[informal]
    \label{thm:GHZ-exc}
    Necessary and sufficient condition for a graph $G$ to be the exclusivity graph of an $n$-context GHZ-type paradox is 
    \begin{align}
        \alpha(G)=n-1, \,\ \vartheta(G)=n, \,\ \mathrm{and} \,\ \chi(\bar G)=n,
    \end{align}
    where $\bar{G}$ is the graph complement of $G$ and $\chi(\bar{G})$ is its chromatic number.
\end{theorem}
Note that the notion of $n$-context throughout this paper refers the number of context-cover instead of the number of context; to complete the logical proof, additional contexts may be counted. The proof of \autoref{thm:GHZ-exc} and the discussion regarding the above point can be found in the \methodsname\ section. As the left-hand side value of Eq.\,(\ref{eq:CSW}) in any no-signaling theory is upper bounded by the exclusivity graph's fractional packing number $\alpha^*(G)$, and $\alpha^*(G)\leqslant\chi(\bar{G})$\,\cite{Acin17}, a quantum correlation associated with a GHZ-type paradox must also be a fully contextual correlation\,\cite{Amselem12} that lays on the boundary of the no-signaling polytope, but the converse is not necessarily true. Our results thus clearly show the role of the GHZ-type paradox as a peculiar class of the strongest quantum correlation allowed by special relativity.

\vspace{10pt}\noindent\textsf{\textbf{A three-context GHZ-type paradox}}\hfill

\begin{figure}[b!]
    \centering
    \includegraphics[width=.99\columnwidth]{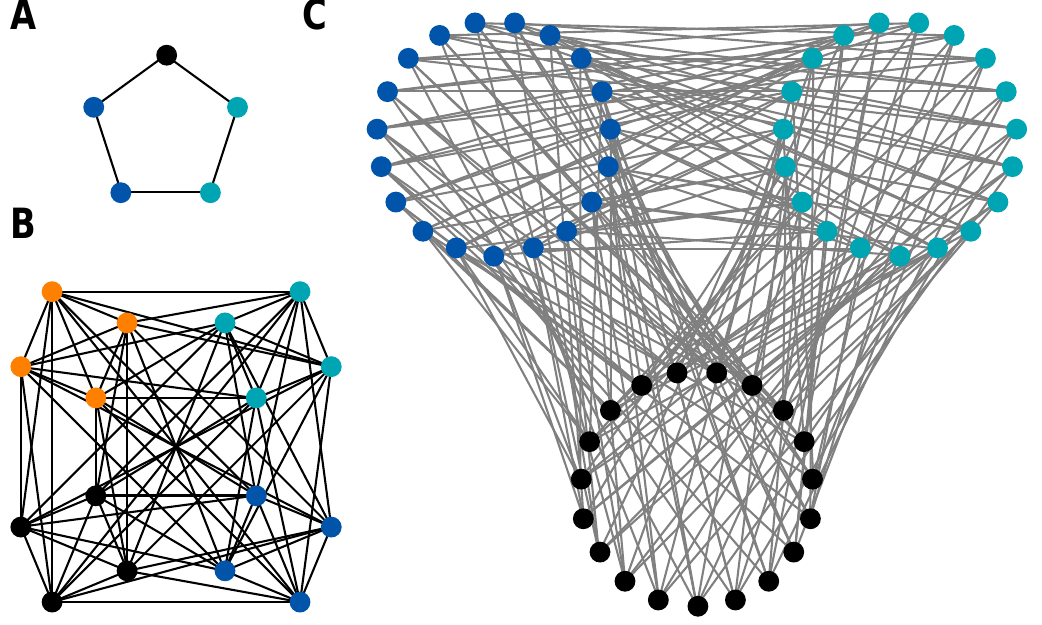}
    \caption{\textbf{Graphs of exclusivity.} The vertices with the same color belong to the same context.  (a) A pentagon is the simplest graph which shows nonclassicality when considered as an exclusivity graph. Measurements with such an exclusivity structure exhibit a three-context Hardy-type paradox, but the quantum success probability is less than 1. (b) The graph complement of the Shrikhande graph is the underlying exclusivity structure of the original GHZ paradox with four contexts. In both (a) and (b) the connected vertices represent mutually exclusive events. (c) The Perkel graph is the orthogonal representation of the rays in the three-context GHZ-type paradox. Note that the exclusivity graph here is complementary to the Perkel graph. Black (gray) lines connect the (non-)exclusive events.
    \label{fig:exc}
    }
\end{figure}

Because the graphs of exclusivity in GHZ-type paradoxes have a fixed $\vartheta-\alpha=1$, a GHZ-type paradox whose events can be covered by fewer contexts is associated with a larger quantum--classical ratio and stronger nonclassicality. However, the minimum of this number is not known: as we will prove in the \methodsname\ section, it must be greater than two, but all the known examples of GHZ-type paradox have the number of context-cover of at least four, leaving the three-context case unexplored. Although noncontextuality inequality\,\cite{KCBS08} and logical proofs of contextuality\,\cite{Cabello13} using three context-cover have been studied in experiments\,\cite{Marques14,Jerger16}, the corresponding exclusivity graph---a pentagon (cf. Fig.\,\ref{fig:exc}(a))---has only a $\vartheta-\alpha=\sqrt{5}-2<1$. As such, the quantum violation is less than $1$ and the logical proof does not constitute a GHZ-type paradox. Further, a GHZ-type paradox can also be constructed from a Kochen--Specker set by selecting one ray as the input state and removing all its orthogonal rays, but all known Kochen--Specker sets employ at least five contexts\,\cite{Mermin93}. The exclusivity graph in the smallest case (cf. Fig.\,\ref{fig:exc}(b)) is exactly the same as that in the original GHZ paradox. The existence of a four-context Kochen--Specker set is thus a long-standing open question. 

All the above observations make the search for a three-context GHZ-type paradox an interesting task. Towards this objective, our next contribution is to identify an exclusivity graph hosting such a paradox.
According to \ref{thm:GHZ-exc}, the desired exclusivity graph $G$ has some fixed graph-theoretic constants, including an independence number of $\alpha(G)=2$ and a Lov\'asz number of $\vartheta(G)=3$; moreover, its graph complement should have a chromatic number of $\chi(\bar G)=3$, that is, $\bar G$ is triangle-free and three-colorable. 
We have identified the graph complement of the Perkel graph\,\cite{Perkel79} as shown in Fig.\,\ref{fig:exc}(c) as a candidate for the exclusivity graph. The approach we use is to search across the graphs with known graph constants, and we do not know if a graph with fewer vertices exists that can satisfy the same conditions, due to the difficulty of traversing the exponentially many possible graphs. Nevertheless, we also prove in the \methodsname\ section that these requirements cannot be satisfied by a strongly regular graph\,\cite{Bose63} which is highly symmetric and widely used in previous proofs of contextuality\,\cite{GHZ89,Cabello96}, thus posing further limitations to the existence of other examples.

Once the candidate exclusivity graph is determined, we can calculate the explicit form of the measurement events by representing them with a set of rays; each ray corresponds to the nondegenerate eigenstate of a projector. These rays form an orthogonal representation of the Perkel graph. The procedure for obtaining the orthogonal representation is to first determine the Gram matrix of the rays by semidefinite programming known as Lov\'asz optimization\,\cite{Lovasz79}, and then solve the individual rays by Cholesky decomposition and Gaussian elimination. The procedure will be extensively discussed in the \smname. We realized the semidefinite programming using a Python package \texttt{cvxopt} and the Cholesky decomposition using \texttt{Mathematica 11.2}. Assuming perfect exclusivity as dictated by the exclusivity graph so the second term of Eq.\,(\ref{eq:CSW}) vanishes, the three-context GHZ-type paradox can be expressed as:
\begin{align}
\begin{alignedat}{3}
    p_1&:={\textstyle\sum_{k=1}^{19}}&&\Pr(1|k)&&=1, \\
    p_2&:={\textstyle\sum_{k=20}^{38}}&&\Pr(1|k)&&=1, \\[3pt] 
    \midrule\midrule
    \quad p_3&:={\textstyle\sum_{k=39}^{57}}&&\Pr(1|k)&&=\Big\{
    \begin{array}{ll}
        0, & \text{NCHV}, \\
        1, & \text{Q}.
    \end{array}
\end{alignedat}
\label{eq:ghz-3ctx}
\end{align}
The explicit definition of the individual rays, as well as the Gram matrix, will be given in the \methodsname\ section. The Gram matrix we found has a rank of 37, which is also the Hilbert space dimension in which the corresponding set of rays can be embedded. As will be further discussed in the \smname, we do not know if a lower-dimensional realization of the Gram matrix exists, as the set of rank-limited matrices is non-convex and we could not efficiently run the optimization.

\vspace{10pt}\noindent\textsf{\textbf{Towards a time-multiplexed optical test}}\hfill

\begin{figure}[t]
    \centering
    \includegraphics[width=.99 \columnwidth]{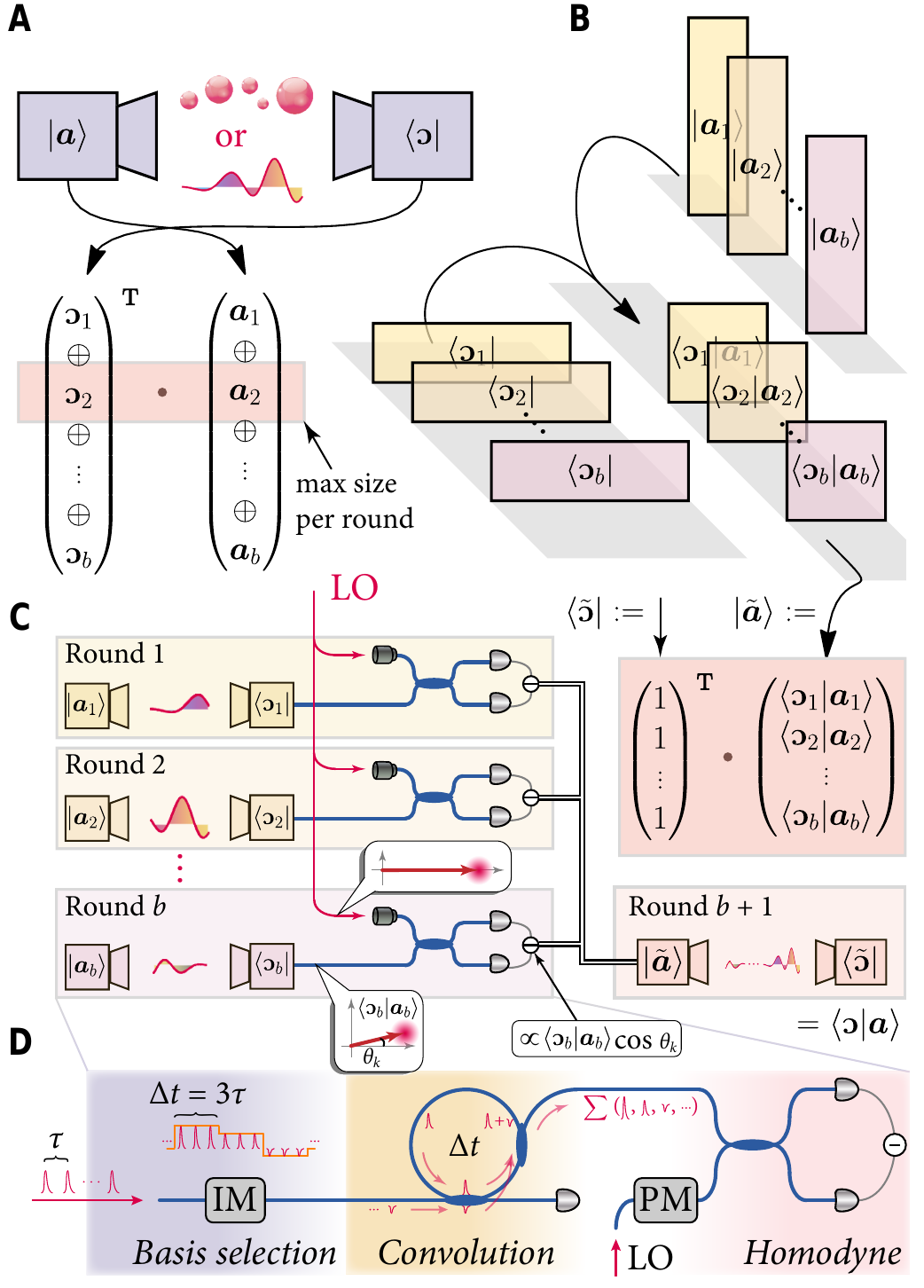}
    \caption{\textbf{Experimental design.} (a) A contextuality test requires a set of prepare-and-measure probabilities, obtained from either single-photon or coherent-state interference. The challenge is the required Hilbert-space dimension can exceed the size that a photonic processor can handle. (b) The high-dimensional inner product can be decomposed into subspaces and evaluated separately. (c) Extraction of the inner product with a local oscillator (LO). In the first $b$ rounds of the experiment, the inner products in subspaces were measured. In the last round, a ray was reconstructed from previous measurement results, and its projection on a unit basis recovered the result of the high-dimensional interference. (d) Sketch of the experimental setup. An intensity modulator prepared various input states. A fiber ring implemented optical convolution. The output mode at a specific time corresponded to the post-measurement state, whose amplitude was subsequently extracted via homodyne detection. IM intensity modulator, PM phase modulator.}
    \label{fig:setup}
\end{figure}

Our proposed three-context GHZ-type paradox would be worth an experimental test, both because it uses the least possible number of orthonormal basis and that a noncontextuality inequality with a high quantum--classical ratio of $3/2$ can be checked via its comprised measurements. According to Eq.\,(\ref{eq:ghz-3ctx}), to observe such a paradox, it is sufficient to measure the projection probability of the initial state on various measurement basis, and additionally to confirm 
the orthogonality of measurements corresponding to mutually exclusive events. The procedure is akin to Cabello's simplified method for testing contextuality using prepare-and-measure experiments\,\cite{Cabello16} with graph-theoretic approach-based inequalities: as shown in Fig.\,\ref{fig:setup}(a) and (b), it has the merit of requiring no sequential, non-demolition measurements. 

However, the requirement of high-dimensional measurements poses a substantial challenge for experimentally observing the three-context GHZ-type paradox.
While such measurements would in principle be amenable in a multi-qubit system available in different platforms\,\cite{ylchi23, mqiao24, zhbao24}, due to the complicated form of the projectors, they will require many entangling operations and suffer from decreased accuracy. Noticing this point, we chose to use the photons as a natural courier for encoding high-dimensional quantum information.
Our endeavor made use of the observation that a strong coherent light can provide a faithful phase reference---often called a ``local oscillator'' in the context of continuous-variable quantum optics, allowing the extraction of a photonic object's full complex amplitude information. With this information, we were able to devise a strategy to decompose a high-dimensional measurement into relatively lower-dimensional subspaces, thus greatly enhancing the scalability of the platform and facilitating the experiment. Despite the highly desirable scalability, our current setup is not fully compatible with a noncontextual theory description; therefore, we could not claim the experiment to be a test of contextuality in a strict sense. However, this can be fixed by changing the last measurement stage back to photodetection, as we will discuss close to the end of the paper.

We utilized the time-bin degree of freedom of light to encode the analogous time-bin qudit and implement the prepare-and-measure experiment. The choice of degree of freedom is due to that it can best accommodate the requirement of a local oscillator. Concretely, we used the following encoding method to map a $d$-dimensional quantum state onto a series of pulsed coherent states:
\begin{align}
    \ket{\boldsymbol{a}}&=(a_1\, a_2\, \cdots\, a_d)^\dagger \nonumber\\ 
    &\leftrightarrow \{\ket{\alpha_1, \Delta t}, \ket{\alpha_2, 2\Delta t}, \cdots, \ket{\alpha_d, d\Delta t}\}.
    \label{eq:map}
\end{align}
The encoded states contain two entries: The first entry, $\alpha_k=\tilde \alpha a_k$ denotes the displacement of the individual coherent states and $\tilde \alpha$ is a constant determined by the intensity of the input coherent state. The second entry specifies the time that it is generated. The encoding is possible with coherent states because when subjected to an interferometer network transformation, the probability distribution of a single photon at the output ports is the same as the intensity distribution if the single photon is replaced by a coherent state\,\cite{Barnett22}. In our experiment, this coherent state train was created by casting intensity modulation on a pulsed laser. 

\begin{figure*}[t]
    \centering
    \includegraphics[width=1.7 \columnwidth]{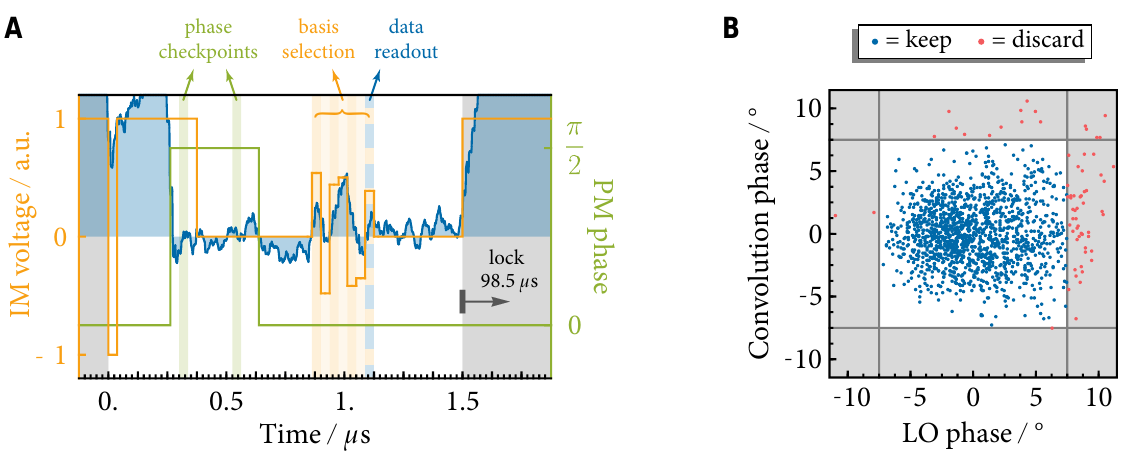}
    \caption{\textbf{Data acquisition.} (a) An exemplary control signal and recorded oscilloscope trace centered at the measure period. The control signal consists of an intensity notch for synchronization, a $\pi/2$-pulse at the local oscillator for phase calibration, and the intensity modulations encoding the preparation and convolution basis. The gray background indicates the lock period and the colored bands highlight the points where operations are performed or data is being taken. (b) The deduced phase error of the convolution fiber ring and the local oscillator. The red-colored data points had at least one phase error greater than $\pi/24$ and were excluded from the final calculation.}
    \label{fig:waveform}
\end{figure*}

To measure the encoded state on a specific basis, hereafter denoted as $\ket{\revc}$, we sent the pulse train into a fiber ring that has a round-trip time of $\Delta t$, so a pulsed coherent state would meet and interfere with another emitted at a later time after it circumnavigated the ring. The output modes from the ring thus acquired a component from every earlier pulse. Effectively, it can be expressed as a discrete convolution of the input state $\boldsymbol{a}*\boldsymbol{c}_0$ with the kernel of convolution being $\boldsymbol{c}_0=\{c_1,c_2,\ldots,c_k,\ldots\}$, where $c_k$ denote the complex amplitude ejection ratio of a pulse upon its $k$-th encounter of the output coupler. Focusing on the output pulse at time $d\Delta t$, we found that it will have an amplitude of $\sum_{k=1}^d a_{(d+1-k)}c_k$ which is equivalent to the inner product, $\braket{\revc_0|\boldsymbol{a}}$, where $\revc_0=\{c_d,c_{d-1},\ldots,c_1\}$ represents the first $d$-terms of the entry-reversed convolution kernel. Further, the basis of the projective measurement $\ket{\revc}$ can be adjusted from $\ket{\revc_0}$ by driving the intensity modulator before the pulses enter the fiber ring. The probability of a qudit measurement can thus be efficiently evaluated by looking at the strength of the output pulse at a specific time.

In practice, a fiber ring is inevitably accompanied by ejection into unwanted time-bins, insertion losses, and chromatic dispersion, so the ejection ratio quickly decreases for a large number of loops. As such, the kernel of convolution only has a limited number of meaningful nonzero terms, hampering the capability of measuring high-dimensional states. To overcome this limitation we notice the input state and measurement basis can be written into a direct-sum expansion: $\boldsymbol{a}=\bigoplus_{k=1}^b\boldsymbol{a}_k, \revc=\bigoplus_{k=1}^b\revc_k$, where $b$ is the total number of partitions, such that ${\rm Dim}(\boldsymbol{a}_k) = {\rm Dim}(\revc_k)$ and $\sum_{k=1}^b {\rm Dim}(\boldsymbol{a}_k) = {\rm Dim}(\boldsymbol{a})$. Then, the inner product can be expressed as:
\begin{align}
    \braket{\revc|\boldsymbol{a}}=\textstyle\sum_{k=1}^b\braket{\revc_k|\boldsymbol{a}_k}.
    \label{eq:ds-expan}
\end{align}
The decomposition thus offered a possibility to divide-and-conquer the high-dimensional interference. The challenge in using such a decomposition is that the summation is taken for complex amplitudes instead of photodetection probabilities, and monitoring the power output for each of the right-hand side terms does not provide sufficient information for reconstructing the projection probabilities.

We employ homodyne detection to resolve this final challenge and provide the setup scalability: by interfering the optical mode with the local oscillator on a balanced beam splitter, the intensity difference between the two output ports will be proportional to the mode's in-phase amplitude with the local oscillator. Since the projectors in the three-context GHZ-type paradox (Eq.\,(\ref{eq:ghz-3ctx})) comprise only real numbers, the in-phase amplitude can already capture all information of the input coherent states after convolution, and thus the complex amplitude was recovered by experimentally observable quantities.

Overall, our experimental design is outlined in Fig.\,\ref{fig:setup}(c); a concrete example of encoding the pulse sequence for a prepare-and-measure experiment is provided in the \smname. Firstly, we expressed the state $\ket{\boldsymbol{a}}$ as the decomposed form: $\boldsymbol{a}=\bigoplus_{k=1}^b\boldsymbol{a}_k$. For each sub-normalized state
$\boldsymbol{a}_k$ we employed the optical convolution with the kernel $\revc_k$ and homodyne detection to register the amplitude on the desired partial measurement basis. Measurement of all the 
$b$ groups of inner products yielded a sequence of $b$ elements---all the right-hand-side terms in Eq.\,(\ref{eq:ds-expan}). To extract the initial high-dimensional inner product, we encoded these $b$ elements as a new ray: $\ket{\tilde{\boldsymbol{a}}} = \{\braket{\revc_k|\boldsymbol{a}_k}\}_{k=1}^{b}$.
By setting the kernel to be $\bra{\tilde{\revc}} = \{1, 1, \ldots, 1\}^{\rm T}$, the initial inner product $\braket{\revc|\boldsymbol{a}} = \braket{\tilde{\revc}|\tilde{\boldsymbol{a}}}$ can be measured via another round of optical convolution. 
Finally, the measurement probabilities in Eq.\,(\ref{eq:ghz-3ctx}) was calculated as the absolute square of the homodyne results normalized against a set of orthonormal basis.

\vspace{10pt}\noindent\textsf{\textbf{Experimental implementation}}\hfill

We realized the prepare-and-measure experiment in a fiber-based, electro-optic modulated photonic processor as shown in Fig.\,\ref{fig:setup}(d); the full experimental setup is extensively described in the \methodsname\ section. We have developed a \texttt{MATLAB 2020b} script to control an arbitrary function generator and a digital oscilloscope to run a fully automated experiment. 

A correct and stable phase at different parts of our setup is a crucial requirement for implementing the desired convolution and homodyne detection. To this objective, we used a photodetector at the second output port of the convolution ring's injection fiber coupler and the DC monitor output of the homodyne detector to extract the phase information, and adopted a lock--measure scheme to alternate the setup between a phase locking operation and the pulse measurement. The details about the individual locks can be found in the \methodsname\ section. The scheme ran on a 10 kHz cycle: in the first 98.5 $\mu$s of the cycle (lock period), the intensity modulator was set at maximal transmission, and the servo (TEM Messtechnik LaseLock) implemented active phase stabilization. In the last 1.5 $\mu$s (measure period), the electro-optic modulators delivered the modulations to prepare the actual input state and implement the change of the measurement basis. The resulted output intensity from the setup was much weaker than in the lock period; however, due to the low-pass filter in the servo and the limited frequency-response range, the piezo controlling the phases of the setup could not respond to the sudden change of the input. This way, the phases would remain almost constant during the measure period.

We registered the homodyne outcome with an oscilloscope. An exemplary waveform together with the curves of modulation voltages is depicted in Fig.\,\ref{fig:waveform}(a). When the last pulse for encoding the input state exited the fiber ring, the homodyne result was recorded, and the outcome would be proportional to one term of the desired inner product (right-hand side of Eq.\,(\ref{eq:ds-expan})). 
To verify the locking phase, we added a $\pi/2$ phase on the local oscillator before the pulses for measurement basis selection, and observed the response from the homodyne detector. As detailed in the \methodsname\ section, both the convolution and the homodyne phases can be extracted from two checkpoint voltages in the waveform. The result of phase calibration is shown in Fig.\,\ref{fig:waveform}(b). The standard deviations of the convolution phase and the local oscillator phase error were measured to be $2.74^\circ$ and $3.94^\circ$. Additionally, we discarded all data points with at least one phase error greater than $7.5^\circ$ to avoid undesired noise.

\begin{figure}[tb]
    \centering
    \includegraphics[width=.99 \columnwidth]{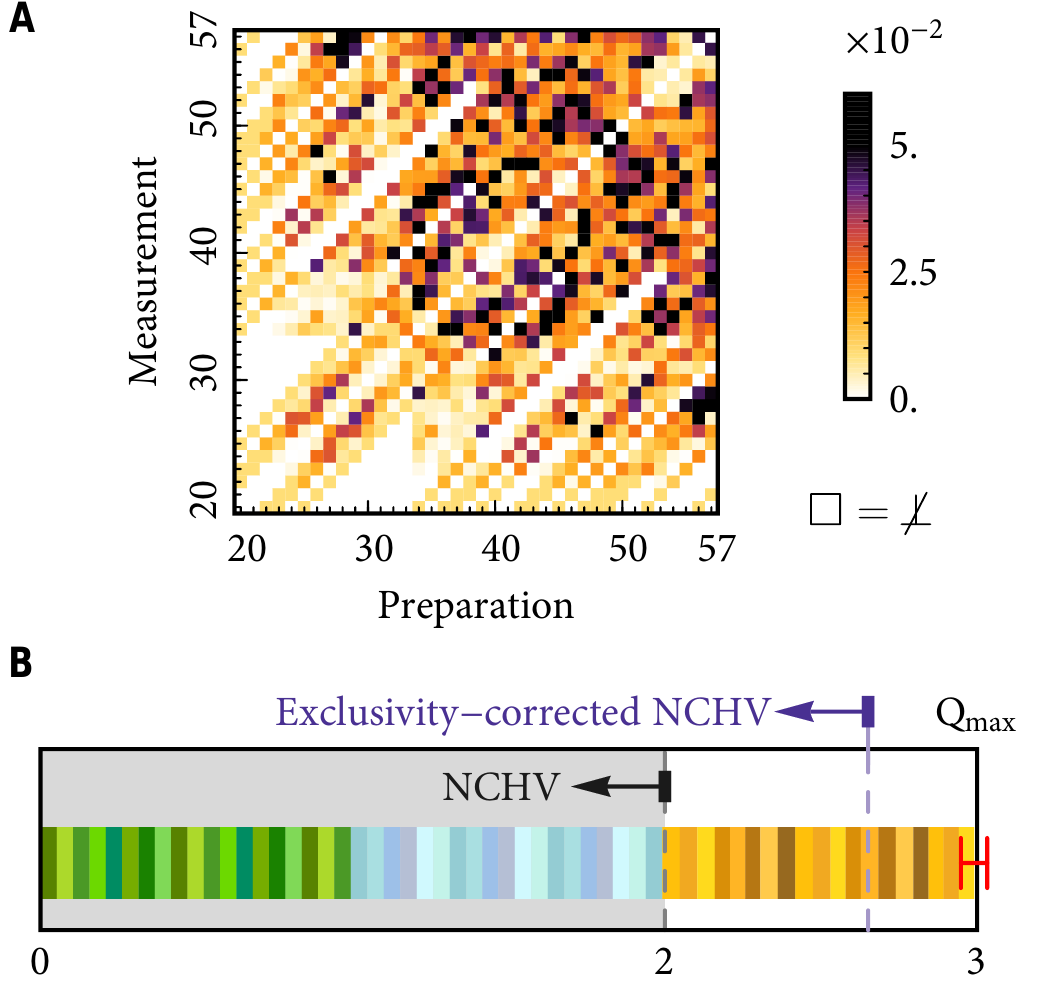}
    \caption{\textbf{Experimental results.} (a) Calculated values of the second term in (\ref{eq:CSW}) for each pair of compatible projectors that does not equal the computational basis. The colors stand for the probabilities of ideally orthogonal measurements not giving exclusive responses; for non-orthogonal measurements rays the corresponding boxes are left as white. (b) Experimental results for the three-context GHZ-type paradox vs. the predictions of noncontextual models, exclusivity-corrected noncontextual models and the quantum theory. The three colored bands correspond to the three sums of probabilities in Eq.\,(\ref{eq:ghz-3ctx}), respectively.
    }
    \label{fig:result}
\end{figure}

With the precise electro-optic modulation and phase-locking, the obtained measurement probabilities closely resembled that would be in an ideal quantum prepare-and-measure experiment. The ideality can be best witnessed in that whenever the state rays of preparation and measurements were orthogonal, the homodyne detection amplitude would almost vanish. We confirmed the above witness by implementing the prepare-and-measure procedure corresponding to the $|E(G)|=1425$ pairs of exclusive rays and projectors in the orthogonal representation of the Perkel graph. Note that the procedures were not implemented if at least one of the prepared or measured rays equaled the computational basis; in which case the probability would only depend on the extinction ratio of intensity modulation and not on interference. The results for the remaining events (cf. Fig.\,\ref{fig:result}(a)) demonstrated an average detection probability for such a procedure of only $1.74(11)\%$. Henceforth, the error bars corresponding to $1\sigma$ standard deviations were obtained from bootstrapping. The high orthogonality ensured the statistics from our prepare-and-measure procedure conformed to the requirements of exclusivity, so the prerequisite of exclusivity in the theory was indeed fulfilled in this platform.

Next, we proceed to test the three-context GHZ-type paradox by directly measuring the three sums of probabilities in Eq.\,(\ref{eq:ghz-3ctx}). Our experimental result gives:
\begin{align}
    p_1=0.9939(15),~ p_2=0.9980(2),~ p_3=0.9983(2).
    \label{eq:ghz-result}
\end{align}
It was in excellent accord with quantum predictions and displayed strong disagreement with the prediction of the noncontextual theories. 

Note that the orthogonality between exclusive events is never perfect in a real experiment, so the observation of a GHZ-type paradox should be taken as a strong, but qualitative indication of contextuality. To substantiate the observation, we further implement a test of the noncontextuality inequality (Eq.\,(\ref{eq:CSW})) associated with the same exclusivity graph. Violation of such an inequality compensates for the deviation from ideal exclusivity and can refute noncontextual models with realistic experimental data\,\cite{Cabello16}. 
The test is made possible using the same data from the GHZ-type paradox which gives the first term of Eq.\,(\ref{eq:CSW}), together with a verification of orthogonality between all projectors supposed to be orthogonal to each other, which gives the first term of Eq.\,(\ref{eq:CSW}). We describe the detailed procedure in the \methodsname\ section. We observe:
\begin{align}
    \sum_{i\in V}~~ \Pr(1|i) =& 2.9902(4), \nonumber\\ ~\alpha(G) + \sum_{(i,j)\in E} \Pr(1,1 &|i,j) = 2.651(4)
    \label{eq:CSW-result}
\end{align}
which means that, after correcting for the imperfect exclusivity and adding the correlation term to the ideal-case classical bound, the experimental data still violated the upper bound of noncontextuality by 8.06 standard deviations and thus rejecting such a description with almost absolute confidence.

\vspace{10pt}\noindent\textsf{\textbf{\large Discussion}}

In conclusion, we have derived and observed a highly exotic quantum correlation by exploring the potential of optical interferometry. Firstly, by developing the graph-theoretical approach for the study of the GHZ-type paradoxes, we have found a strong GHZ-type contextuality whose constituent events can be included in only three contexts---the least possible number of contexts in theory, but still reveal a deterministic contradiction between classical and quantum descriptions of the same underlying correlations. When translated into experimentally testable noncontextuality inequality, the argument also gives the largest quantum--classical ratio among all GHZ-type paradoxes. Interestingly, the argument was found by examining named graphs with known graph-theoretic constants, instead of using the modern computer-assisted search method\,\cite{Cabello15,zpxu20peres} or being derived from a Bell inequality already with a large quantum violation\,\cite{zhliu23phcp}. In fact, with a vertex clique cover number less than four, the exclusivity graph here cannot host any Bell-type inequalities. Our result thus highlights the link between the most exotic quantum correlations and graphs with high degrees of symmetry and may shed light on the search for other strong forms of quantum correlations. 

Secondly, we have exploited a time-multiplexed optical setup to reproduce the full probability distribution of the three-context GHZ-type paradox in an analogous system. By virtue of the direct sum decomposition of a high-dimensional state into lower-dimensional components and the detection of the full complex amplitude information using the homodyne detection, we dramatically scaled up the dimensionality that the optical interferometry can study and implemented a 37-dimensional prepare-and-measure experiment with a precision high enough to demonstrate contextuality. 
Our result clearly showcased the power of combining the time-bin interferometry and homodyne detection, and it would not be hard to directly apply this method in the study of synthetic dimensions of photons\,\cite{lqyuan18,Weidemann20,Leefmans22} to reveal the rich physics in phase space and more complicated quantum dynamic features in high-dimensional systems.

Our experimental platform using the time-multiplexing technique is also well-compatible with quantum computing applications. The technique was adopted in large-scale entanglement (i.e., cluster state) generation\,\cite{Furusawa13} and Gaussian operations based on measurement-based quantum computing\,\cite{Larsen21}; moreover, non-Gaussian operations can also be implemented with homodyne measurements together with photodetection\,\cite{Sakaguchi23} to obtain universality. Another relevant application would be the Gaussian boson sampling\,\cite{Hamilton17} that is a leading approach to quantum computational advantage\,\cite{Xanadu22}. When photons from multiple output times are looked at, the convolution ring setup is equivalent to the fully connected interferometer in Reference \cite{Sempere22} and can gain universality by adding another layer of ring\,\cite{Motes14, yhe17,syu23,yonezu23}. Compared to the interferometer based on the spatial degree of freedom, the platform here has the advantage of being relatively easy to build, economically friendly, and possessing better scalability.


A limitation of the current experiment is that the setup does not include any kind of operation that produces discretized event outcomes. This causes us to be unable to detect contextuality in the normal sense by measuring the probability of individual events. We could recover the probability distribution that violates the noncontextuality inequality, but these probabilities cannot be interpreted in the way that a discrete-variable noncontextuality theory would require. In fact, as the Wigner function of the system is nonnegative, it can be taken as a continuous-variable hidden-variable description and it is noncontextual. Also, the current process of probability calculation involves Eq.\,(\ref{eq:ds-expan}) which in turn requires quantum theory. However, we believe both of the points can be fixed by using photodetection as the last stage of the measurement. In \smname, we describe a setup that would address all the above issues and promote the experiment to a real contextuality test. It would be a reasonable to target at realizing such a setup in a succeeding work.

We believe this work has opened several avenues for future research. From a fundamental perspective, the existence of a three-context GHZ-type paradox leaves open the possibility of finding a Kochen--Specker set whose events can be covered by four contexts. It would be interesting to consider if such a set can already be constructed as a state-independent version of the GHZ-type paradox here, using some methods of expansion, e.g., as in Reference\,\cite{Cabello21}. Pertinent to this topic, we also prove in the \methodsname\ section that if any four-context Kochen--Specker set does exist then the four contexts must be disjoint; hence, this Kochen--Specker set would not accept a simple parity proof\,\cite{Cabello96,Lisonek14}. Finally, an interesting endeavor would be to convert the large amount of nonclassicality in the three-context GHZ-paradox to a quantum advantage. Based on contextuality, quantum advantage in shallow circuits has been composed\,\cite{Bravyi20}. We hope our findings can be used to build even stronger quantum advantages in high-dimensional systems.

\vspace{10pt}\noindent\textsf{\textbf{\large\methodsname}}\hfill

\begin{figure*}[t]
    \centering
    \includegraphics[width=.99\textwidth]{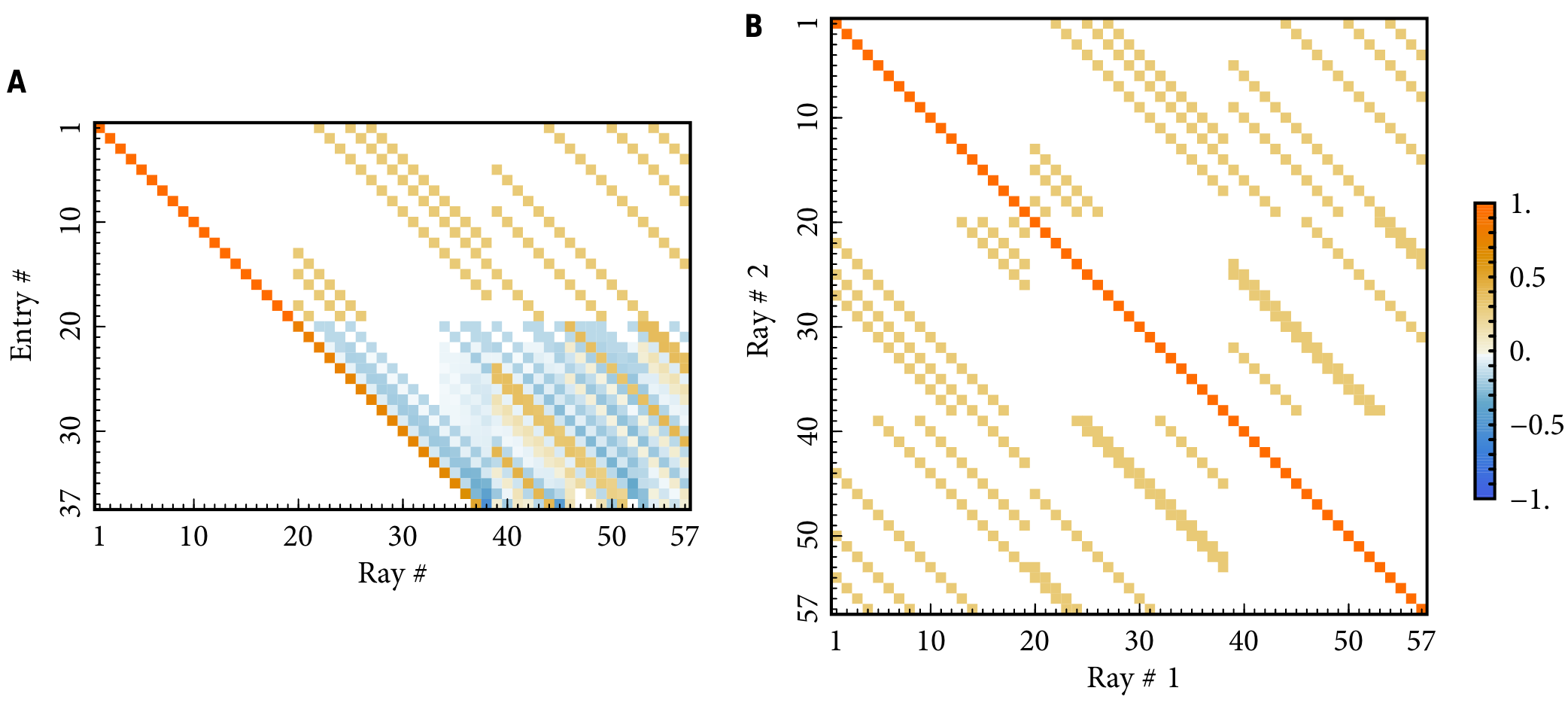}
    \caption{\textbf{Numeric settings.} (a) An orthogonal representation of the Perkel graph obtained from Lov\'asz semidefinite programming and Cholesky decomposition. (b) The Gram matrix of the orthogonal representation of the Perkel graph. The values of the diagonal (red-ish), the yellow-ish and the white entries are 1, 1/3, and 0, respectively.}
    \label{fig:ext-gmats}
\end{figure*}

\noindent\textbf{Proofs of the propositions.} Here we prove the necessary and sufficient conditions for a graph to be the exclusivity graph of a set of rays in a GHZ-type paradox (Theorem 1). This gives a corollary that the minimum number of context-cover in a GHZ-type paradox is three (Theorem 2). Then, we show that no strongly regular graph can host a three-context GHZ-type paradox (Theorem 3), and that if a Kochen--Specker set consists only four context-cover, then those contexts are all disjoint (Theorem 4). Finally, in Fig.\,\ref{fig:ext-gmats}, we give the explicit definition of the individual rays in the three-context GHZ-type paradox and the corresponding Gram matrix.
We restate every proposition before the proof to keep the results self-contained.

\vspace{7pt}\noindent\textit{Theorem 1. }{\it
    A graph $G$ is the exclusivity graph of an $n$-context GHZ-type paradox iff. 
    $ \alpha(G)=n-1, \,\vartheta(G)=n,$ and $\chi(\bar G)=n,$
    where $\bar{G}$ is the graph complement of $G$. $\alpha(G), \,\vartheta(G),$ and $\chi(G)$ are $G$'s independence number, Lov\'asz number, and chromatic number, respectively.}

\begin{proof}
    Necessity---The first two equations can be justified by adding all the conditional probabilities in Eq.\,(\ref{eq:ghz-general}) together, comparing the sum with Eq.\,(\ref{eq:CSW}) and noticing that the second term vanishes for ideal exclusivity. To prove the last equation, we observe that $\bar G$ can be colored by assigning the same color to all vertices that represent the events in the same context and using a different color for each context. This explicit construction guarantees $\chi(\bar G)\leqslant n$. But according to Lov\'asz's sandwich theorem\,\cite{Grotschel81}, $\chi(\bar G)\geqslant\vartheta(G)=n$, so it must be $\chi(\bar G)=n$.

    Sufficiency---The classical bound can be trivially proved using the original graph-theoretic approach to contextuality\,\cite{CSW14}. To prove the quantum part that the event probabilities can be disassembled into groups of unit probabilities given the graph constants, we notice that the Lov\'asz number of a graph $G$ is the optimum of the objective function in the following semidefinite program (``Lov\'asz optimization''):
    \begin{alignat}{3}
    {\max_\mathbf{B}}&\quad\vartheta=\tr(\mathbf{BJ}) \label{eq:lovasz-sdp}\\
    {\rm subject~to}&\quad\mathbf{B} \succcurlyeq 0, \tr(\mathbf{B}) = 1, \nonumber\\
    &\quad B_{ij} = 0,\;\forall\ (i,j)\in E(G), \nonumber
\end{alignat}
where $\mathbf{J}$ is an $N\times N$ matrix with all of its entries being 1, $(\cdot)\succcurlyeq0$ means the matrix on the left-hand side of the operator is positive semidefinite, and $E(G)$ is the edge set of $G$. Let the matrix $\mathbf{B}$ be the Gram matrix of the set of rays $\boldsymbol{r}_k, k\in[1, N]$ and denote $G'$ as the exclusivity graph of $\ket{\boldsymbol{r}_k}\bra{\boldsymbol{r}_k}$. As the last constraint in Eq.\,(\ref{eq:lovasz-sdp}) guarantees $\braket{\boldsymbol{r}_i|\boldsymbol{r}_j}=0, \,\forall\ (i,j)\in E(G)$, it must be $E(G)\subseteq E(G')$ and thus $E(\bar{G'})\subseteq E(\bar{G})$, therefore, a $\chi(\bar{G})$-coloring for $\bar{G}$ is also a proper coloring for $\bar{G'}$. We denote such a coloring as $V(\bar G)\to[j, k]$, where $j$ indicates the color and $k\in[1, m_j]$ is the index of the vertex among those $m_j$-vertices with the same color. Now, the vertices with the same color in $\bar{G}$ must correspond to mutually exclusive events in $G'$. Due to the exclusivity principle\,\cite{Cabello13KCBS}, we have $\sum_{k=1}^{m_j}\Pr(1|[j, k])\leqslant1, \forall j$, but from the definition of the Lov\'asz number we also have $\sum_{j=1}^{\chi(\bar{G})}\sum_{k=1}^{m_j}\Pr(1|[j, k])=\chi(\bar G)$. Therefore, it must be $\sum_{k=1}^{m_j}\Pr(1|[j, k])=1$ for all $j$. The GHZ-type paradox in Eq.\,(\ref{eq:ghz-general}) is thus recovered.
\end{proof}

\vspace{7pt}\noindent\textit{Theorem 2. The lower bound of the number of context-cover in a GHZ-type paradox is three. There exists no GHZ-type paradox using one or two context-cover.} 

\begin{proof}
    By Theorem 1, the underlying graph $G$ of exclusivity corresponding to an $n$-context GHZ-type contextuality has an independence number of $\alpha(G)=n-1$. An $n=1$ paradox would require $\alpha(G)=0$ which is a null graph: $|V(G)|=0$, where the measurement events are undefined. An $n=2$ paradox would require $\alpha(G)=1$ which is a complete graph, but a complete graph also has a Lov\'asz number of $\vartheta(G)=1=\alpha(G)$, and thus its corresponding noncontextuality inequality cannot have any quantum violation.
\end{proof}

\vspace{7pt}\noindent\textit{Theorem 3. No strongly regular graph hosts a three-context GHZ-type paradox.} 

\begin{proof}
    We sketch the proof in this section and defer the complete proof to the \smname. A strongly regular graph with four parameters $n,k,a$ and $c,$ denoted by ${\rm SRG}(n,k,a,c),$ is a $n$-vertex $k$-regular graph such that every two adjacent vertices have $a$ common neighbors, and that every two non-adjacent vertices have $c$ common neighbors. By Theorem 1, we only need to prove the nonexistence of a strongly regular graph $G$ with a clique number of two and a chromatic number of three such that the Lov\'asz number of its complement graph is also three. 
    
    The proof is by exhaustion. As the exclusivity graph for a three-context GHZ-type paradox is triangle-free, we have $a=0$. Starting from this observation, we can study the spectrum of the adjacency matrix of $G$, and find its only three nonzero eigenvalues are $k$ and $\left(\pm\sqrt{c^2+4(k-c)}-c\right)/2$. Using the Hoffman--Delsarte eigenvalue bound\,\cite{Cvetkovic80,Delsarte73}, we link the eigenvalues to the known graph constants to have $c\leqslant k\leqslant 2c+3$; the condition can be further narrowed to $k\in\{c,2c+1\}$ by looking at the multiplicities of the eigenvalues. Finally, for both of the cases, we show that the resulted graph cannot satisfy the desired graph-constant constraints.
\end{proof}

\vspace{7pt}\noindent\textit{Theorem 4. If there exists a Kochen--Specker set consisting of only four complete contexts, then those contexts are all disjoint.} 

\begin{proof}
    The proof is by contradiction. Let the four complete contexts be $\{C_i\}_{i=1}^4$. Without loss of generality, we can assume $C_1 \cap C_2 \neq \emptyset$ and $v_0 \in C_1 \cap C_2$. According to the relation between GHZ-type proof and Kochen--Specker set~\cite{zpxu20peres}, we can obtain the exclusivity graph $G$ of a GHZ-type proof by removing $v_0$ and other vertices connected to $v_0$, especially all other vertices in $C_1$ and $C_2$. Consequently, this exclusivity graph $G$ can be covered by two cliques, which implies that $\vartheta(G) \le 2$. Since it stands for a GHZ-type proof, $\alpha(G)$ should be strictly less than $\vartheta(G)$, thus the only possibility is that $\alpha(G)=1$. This can only happen in the case that the graph can be covered by one clique, where $\vartheta(G) = \alpha(G) = 1$. This contradicts the assumption that $G$ is the exclusivity graph of a GHZ-type proof.
\end{proof}

\vspace{10pt}\noindent\textbf{Number of context and context-cover.} In calculating the number of contexts in a GHZ-type paradox, two definitions are possible. The number of context is the number of all orthonormal basis that must be used to derive the deterministic contradiction between the noncontextual and quantum theory\,\cite{Lisonek14, Budroni21}. Here, we define the number of context-cover as the number of contexts required to include all the elementary events in a GHZ-type paradox. As we have discussed before, the number of context-cover determines the quantum-classical ratio.

In the graph-theoretic approach, the number of context and context-cover and are naturally linked to the graph constants of the exclusivity graph, $G$, corresponding to the events in a GHZ-type paradox. Note that $E(G)$ only includes the exclusivity relations that are necessary for deriving the logical paradox and excludes the orthogonal relations that ``accidentally'' appear at finding the numeric solution of the exclusivity graph. In Theorem \ref{thm:GHZ-exc}, we have proved that the number of context-cover is the (vertex) chromatic number of $\bar{G}$ or the vertex clique cover number of $G$, denoted by ${\sf vcc}(G)$. In the same vein, the number of context is the edge chromatic number of $\bar{G}$ or the edge clique number of $G$, denoted by ${\sf ecc}(G)$. As for any graph ${\sf ecc}(G) \geqslant {\sf vcc}(G)$ holds, the number of context is at least not less than the number of context-cover.

According to Vorob'ev's theorem, a necessary condition for the existence of a quantum realization producing contextuality is that the graph of compatibility of the observables has cycles of length more than $3$\,\cite{zpxu19, Budroni21}. Therefore, the minimal number of proof-context in a GHZ-type paradox is at least four, which is saturated by e.g., GHZ and Mermin's construction\,\cite{Mermin90GHZ}. In our case, as the Perkel graph has an edge chromatic number of seven, the proof of its corresponding GHZ-type paradox must use at least seven contexts. 

\begin{figure}[bt]
    \centering
    \includegraphics[width=0.98\columnwidth]{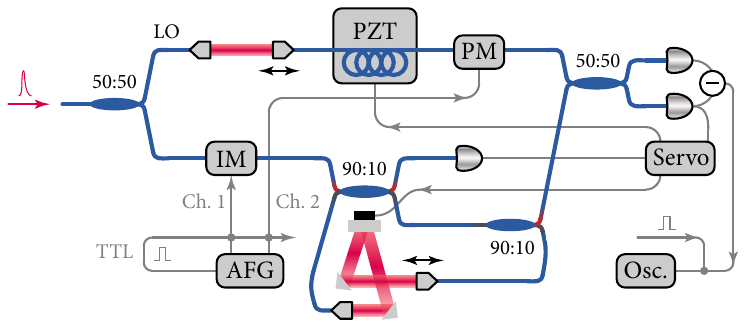}
    \caption{\textbf{Detailed experimental setup.} The blue lines denote optical fibers, and the red strokes indicate light propagating in free-space sessions. Electronic connections are denoted by gray lines. The beam-splitter ports marked with the same color have higher transmissivity. Acronyms: AFG arbitrary function generator, IM intensity modulator, Osc. oscilloscope, PM phase modulator, PZT piezoelectric fiber stretcher.}
    \label{fig:ext-setup-full}
\end{figure}

\vspace{10pt}\noindent\textbf{Experimental setup.} The full experimental setup is shown in Fig.\,\ref{fig:ext-setup-full}. A frequency-locked pulsed fiber laser with a repetition rate of $1/\tau=75.91$ MHz, a central wavelength of 1560 nm, and a spectral bandwidth of 12 nm was adopted as the source of pulsed coherent states. From the parameters of the laser, we estimate the maximal displacement of the coherent state as $\tilde\alpha=1.014\times10^4$, so the shot noise from the fluctuation of photon numbers between individual pulses is comparatively small. 

The 37-dimensional state was encoded in six direct-sum subspaces and implemented in different runs of the experiment; each of the subspaces was in turn represented by up to seven pulsed coherent states. To implement both the preparation of the initial state and the adjustment of the convolution basis, we utilized a commercial lithium niobate intensity modulator (Thorlabs LN81S-FC) with a power extinction ratio of 28 dB. The intensity modulator only generated preparation and measurement basis with real coefficients, but it can be easily upgraded to realize complex coefficients by employing an additional phase modulator. 
The intensity modulator was driven by an arbitrary function generator which has a rising and falling edge time measured as 7 ns. Due to the limited bandwidth of the function generator, we could not address the individual pulses from the laser. Instead, we always modulated three consecutive pulses and only measured the amplitude of the middle one to avoid clashing with the modulation edges; this resulted in an effective pulse frequency of $1/\Delta t=1/(3\tau)$. 

\begin{figure}[tbh]
    \centering
    \includegraphics[width=.72\columnwidth]{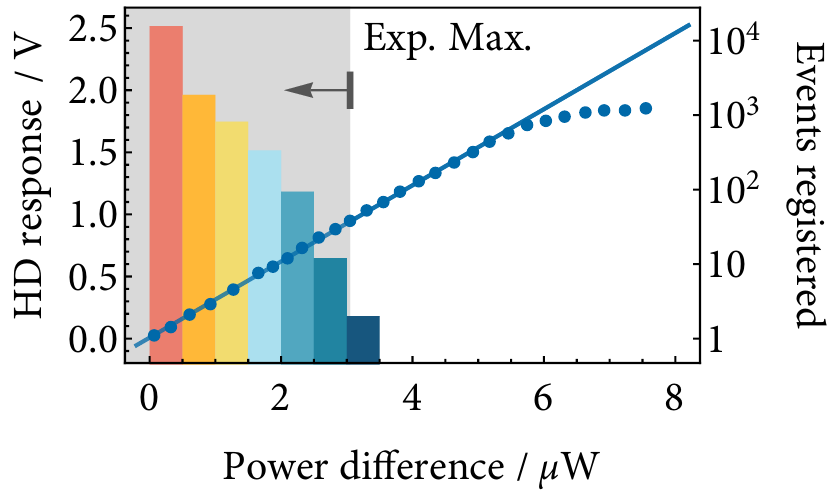}
    \caption{\textbf{Detector response curve.} Data points: homodyne detector (HD)'s voltage response to the power difference between two input ports. Curve: linear fit of those data points with power differences below 5 $\mu$W; the 1 dB gain compression point is at 6.63 $\mu$W. Histogram: distribution of the power differences of the data points registered during the experiment; the maximal power difference was 3.05 $\mu$W and all data points fell in the linear response range.}
    \label{fig:ext-bhd-response}
\end{figure}

To construct the fiber ring for convolution, we utilized two fiber beamsplitters with an amplitude splitting ratio of 90:10 and a free-space delay line. The length of the entire delay line was carefully adjusted to $3\tau$ to align the pulses at different times and maximize the visibility of the interference. We manually calibrated the kernel of convolution instead of using the round-trip loss of the fiber ring cavity (estimated to be 27\%) as the pulses emitted from the ring after different numbers of circulation would have different spectral distribution and interference visibility with the local oscillator. The homodyne detector used in our experiment (Thorlabs PDB480C-AC) had a linear response to power difference up to 5 $\mu$W as shown in Fig.\,\ref{fig:ext-bhd-response}, and all experimental points fell well in the linear regime.

\vspace{10pt}\noindent\textbf{Phase stabilization.} The experimental setup contained two interference points that needed to be phase stabilized. The first is the convolution ring which is a fiber cavity and we adopted the Pound--Drever--Hall technique\,\cite{Black00} to stabilize its length. Concretely, A 3.2 kHz sine-wave modulation signal is sent from the servo to a piezoelectric crystal attached to a mirror in the free-space delay line. The voltage would dither the convolution phase by $\pm 3^\circ$ maximum. As shown in Fig.\,\ref{fig:setup}, we monitored the optical power at one port of the input coupler of the convolution ring. Essentially, this is the ``reflection'' from the fiber cavity and it will be minimized when the convolution phase is an integer multiply of $2\pi$. By demodulating the monitor signal, the servo would be able to find out the direction of the phase drift, and then send a voltage to the piezo to correct the drift and establish the phase lock. 

The second phase lock is required to align the phase of the local oscillator with the rest of the setup. With our lock--measure cycle, this can be ensured by maximizing the power difference at the local oscillator during the lock stage. To this purpose, a dither signal would again be needed to reveal the direction of the phase drift. We sent the 7.2 kHz sine-wave modulation to a fiber stretcher (Optiphase PZ2) in the local oscillator path. The choice of the frequency was to both accommodate the intrinsic frequency response of the fiber stretcher and avoid the high-order harmonic from the upstream lock's dithering. After demodulating and low-pass filtering the signal from the monitor port of the homodyne detector, the resulted error signal would be proportional to the quadrature amplitude (i.e., out-of-phase part to the local oscillator) of the setup, and the servo can drive the fiber stretcher accordingly to stabilize the phase of the local oscillator.

\vspace{10pt}\noindent\textbf{Phase error readout.} As elaborated in the main text, both the convolution and the homodyne phase will remain almost constant during the entire measure period; we denote them as $\varphi$ and $\phi$, respectively. To extract the phase errors, we also need to know the entire kernel of convolution $\{c_1,c_2,\ldots,c_k,\ldots\}$. We pre-calibrated up to 12 terms before the experiment and found the effect of the other terms was sufficiently small. During the locking stage, the readout value of the homodyne detector can be expressed as:
\begin{align}
    h_0 = \tilde\alpha {\rm Re}({\rm e}^{-i\phi} \textstyle\sum_{k=1} c_k {\rm e}^{ik\varphi}).
    \label{eq:phase0}
\end{align}

The phase error readout procedure worked as follows. Firstly, from the state of the lock stage, we added a $\pi/2$-phase to the local oscillator through the phase modulator. When the modulated pulse arrived at the homodyne detector (first green band in Fig.\,\ref{fig:waveform}(a)), the homodyne value became:
\begin{align}
    h_1 = \tilde\alpha {\rm Re}({\rm e}^{-i\phi+\pi/2} \textstyle\sum_{k=1} c_k {\rm e}^{ik\varphi}).
\end{align}
Secondly, we kept the state of the phase modulator and shut off the intensity modulator. This caused no light to enter the fiber cavity and the existing pulses started to leak out of the convolution ring. After three round-trip time (second green band in Fig.\,\ref{fig:waveform}(a)), the homodyne value became:
\begin{align}
    h_2 = \tilde\alpha {\rm Re}({\rm e}^{-i\phi+\pi/2} \textstyle\sum_{k=4} c_k {\rm e}^{ik\varphi}).
\end{align}
Comparing the values of $h_0, h_1$, and $h_2$ gives the desired phase errors $\varphi$ and $\phi$.

\vspace{10pt}\noindent\textbf{Correcting for non-ideal exclusivity.} To measure the correlator terms in Eq.\,\ref{eq:CSW} and arrive at Eq.\,\ref{eq:CSW-result}, we notice that the term $P(1,1|i,j)$ as the probability of the measurement outcome on the first projector $i$ is $+1$, and then the measurement outcome of the post-measurement state, which is the $+1$-eigenstate of the projector $i$, on the second projector $j$ is again $+1$. Experimentally, we prepare the $+1$-eigenstate of the projector $i$, and measure it with the other projectors that are supposed to be orthogonal to it. Define the probability of finding the $+1$-eigenstate of the $i$-th projector on the $+1$-eigenstate of the $j$-th projector to be $P(1|j, i=1)$, we have:
\begin{align}
    P(1,1|i,j) = P(1|i)P(1|j, i=1).
\end{align}
By iterating the procedure over all the possible combinations $(i, j)\in E(G)$, we obtained all the probabilities $P(1|j, i=1)$ as shown in Fig.\,\ref{fig:result}(a). The probabilities $P(1|j, i=1)$ have already been recorded in the observation of the GHZ-type paradox. From these probabilities, we can calculate the desired correction due to the non-ideal exclusivity.

\vspace{10pt}\noindent\textbf{Error analysis.} We identify four main sources of the experimental error. The first was the dither for the phase lock. To lock at the top of an interference fringe generally necessitates a dither signal and demodulation, but the dither also disturbs the lock itself. In our experiment, the inferred level of phase noise was very close to the dither signal, and we could not further weaken the dither and still have a stable lock. This could be fixed by only sending the dither signal during the lock stage and freezing the output of the servo; such an operation, although cannot be incorporated in our current setup, has native support in some software\,\cite{Pyrpl23} that implements versatile field-programmable gate array-based phase locks. 

The second was the imperfection of the intensity modulator, which has both a limited extinction ratio and a drift of zero-output voltage due to the heating of the element. We strived to eliminate the drift by sending the pulses used in the experiment for one hour before data registration so the element was sufficiently ``thermalized'' with the environment. This could be further improved by cascading more intensity modulators to increase the extinction ratio and introducing a feedback control loop to auto-correct the zero-output voltage.

The third was the imperfect interference visibility. The laser used in the experiment was a femtosecond pulsed laser and the pulses were very localized in space. After the spectral filtering, we could only observe an interference when the distance of the two pulses was below 0.4 cm. This made the adjustment of the fiber ring's length challenging. Moreover, the chromatic dispersion in the fiber ring system causes the pulses traveled different times inside the ring to have different spectra. As such, even when a specific preparation and measurement setting should result in perfect destructive interference, the output from the fiber ring could not be eliminated and would be seen by the homodyne detector. We consider this effect as the culprit for the imperfect exclusivity in Fig.\,\ref{fig:result}(a).

Finally, that we cannot fully exploit the 10-bit readout precision of the oscilloscope also limited the precision of the experiment. Due to the phase error readout scheme, the entire oscilloscope trace must not go out of range during the lock period (otherwise $h_0$ in Eq.\,(\ref{eq:phase0}) cannot be measured). This prevented us from selecting a smaller voltage scale and achieving better readout resolution. In the future, it would be helpful to put the phase error estimation subroutine to the servo and use the full precision of the oscilloscope to register the data.

\let\oldaddcontentsline\addcontentsline
\renewcommand{\addcontentsline}[3]{}

\vspace{10pt}\noindent\textsf{\textbf{Acknowledgements.}}---We would like to thank Ad\'{a}n\ Cabello, Jonatan\ B.\ Brask, Daniel\ Cavalcanti and Jonas\ S.\ Neergaard-Nielsen for inspiring discussion and Lucas~Nunes~Faria for helpful feedback on an earlier version of the paper.

\bibliographystyle{rev4-2prx}
\bibliography{tcCTX.bib}

\begin{thebibliography}{87}
\bibitem [{not()}]{note1}%
  \BibitemOpen
  \href@noop {} {}\bibinfo {note} {Throughout this paper, we omit the qualifier
  ``Kochen--Specker'' or ``quantum'' before ``contextuality'' for the sake of
  brevity.}\BibitemShut {Stop}%
\end{thebibliography}

\let\addcontentsline\oldaddcontentsline
%


\appendix
\onecolumngrid
\bigskip\bigskip

\setcounter{equation}{0}
\setcounter{figure}{0}
\renewcommand{\theequation}{S\arabic{equation}}
\renewcommand{\thefigure}{S\arabic{figure}}


\begin{center}
\textbf{\large Supplementary Material for ``Exploring the boundary of quantum correlations \\ with a time-domain optical processor''}
\end{center}

\tableofcontents

\subsection{GHZ-type paradoxes for quantum computing}

It is known that the Clifford (Gaussian) subtheory in discrete- (continuous-)variable quantum theory is efficiently simulable with a classical computer\,\cite{Gottesman98, Braunstein05} and cannot provide acceleration for quantum computing. Starting from this observation, contextuality has been flagged as crucial for quantum computing through its link to negativity \,\cite{Delfosse17, Booth22}. On the other hand, although it is also known that the simulation of contextual correlations is hard (e.g., by requiring a quadratic cost of memory\,\cite{Kleinmann11, Larsson22}), how to derive practical acceleration from contextual relation is a more subtle topic. In this regard, the Greenberg--Horne--Zeilinger (GHZ)-type contextuality is a noteworthy resource. Here, we discuss how the ``perfect correlations'' in the GHZ-type contextuality enable the three different approaches to quantum computing;  for an extended review of the role of contextuality (not necessarily GHZ-type) in quantum computing we would point the readers to the Section VIA of Reference \cite{Budroni21}.

\begin{itemize}
    \item \textit{GHZ-type contextuality signifies maximal magic.}---Howard et al.\,\cite{Howard14} identified contextuality as the necessary condition for a quantum state to be useful in the ``magic state distillation'' subroutine. which enables universal quantum computation. In this paper, the constructed noncontextuality inequality also comes from the graph-theoretic approach. Moreover, for an odd prime dimension $p$, the exclusivity graph $G_p$ satisfies $\alpha(G_p)=p^3$, $\vartheta(G_p)=\chi(\overline{G_p})=p^3+1$, where $\bar G$ means the graph complement of $G$. From the Theorem 1 in the main text, we know that $G_p$ is the exclusivity graph of a $(p^3+1)$-context GHZ-type contextuality: when a quantum state is the maximal resource state for magic state distillation, it will demonstrate a GHZ-type contextuality using Howard et al.'s projector construction. 

    \item \textit{GHZ-type contextuality promotes measurement-based quantum computation.}---Using the sheaf-theoretic approach to contextuality, Abramsky et al.\,\cite{Abramsky17} found that when one uses an empirical model with a contextual fraction of $\lambda$ for an $l2$-measurement-based quantum computing task, the success probability of the task will be upper bounded by $\lambda$. The contextual fraction of an empirical model is the infimum of the weight of the contextual part when the model is decomposed into a noncontextual model and a contextual, no-signaling model. For a no-signaling model, the sum of the event probabilities is upper bounded by the fractional packing number $\alpha^*(G)$ of the exclusivity graph $G$ corresponding to the set of events. In the case of GHZ-type contextuality, the sum of the event probabilities is $\vartheta(G)=\chi(\bar G)$ and as, from graph theory, we know that $\chi(\bar G)\geqslant\alpha^*(G)$, we are able to conclude that the empirical model in a GHZ-type contextuality has a contextual fraction of 1. Therefore, the model will be an ideal resource for measurement-based quantum computation. The application has been established in some small-scale cases. For example, a tripartite GHZ-type correlation can enable the deterministic computation of an \textsc{nand} gate\,\cite{Anders09}. Such a correlation can promote an oracle only able to implement parity check to a universal computer and is thus of practical relevance.

    \item \textit{GHZ-type contextuality brings shallow-circuit quantum advantage.}---The quantum advantage in shallow circuits refers to a type of computational problem, that is solvable by a constant-depth quantum circuit with high certainty, whereas any classical circuit solving the problem with the same certainty must have a depth growing logarithmically with the input size. Such a quantum advantage does not rely on the polynomial hierarchy in the computational complexity theory; it is known to exist\,\cite{Bravyi18} and can tolerate a small amount of noise\,\cite{Bravyi20}. At the center of this kind of quantum advantage is the GHZ-type contextuality: the quantum circuit can solve the problem due to its ability to generate ``perfectly'' contextual correlations. The contextual correlation must be GHZ-type in order to guarantee the output is deterministic regardless of the input size; a Bell-type correlation that only statistically outperforms a local hidden-variable model will fail to do so. This way, the GHZ-type contextual correlation levitated the memory or communication cost in the classical circuit which finally converted into the penalty in circuit depth. Also, the GHZ-type contextuality is necessary to bring the error rate of the circuit below the fault-tolerance threshold and make the quantum advantage robust.
\end{itemize}
All the above examples unravel a strong link between GHZ-type contextuality and quantum computing. Remarkably, at least two of these approaches (i.e., measurement-based quantum computing and quantum advantage in shallow circuits) require a GHZ-type contextuality beyond a Bell-type correlation. In this regard, the Theorem 1 in the main text capable of identifying GHZ-type contextuality from the exclusivity graph of measurements paves a step toward these applications. Also, the 37-dimensional GHZ-type contextuality has a large quantum--classical ratio of $3/2$ that exceeds all the known previous examples for a single quantum system (see, for example, Figure S2 in the supplemental material of Reference \,\cite{zhliu23phcp} for a comparison). We envisage that a shallow quantum circuit can be constructed based on the contextuality identified here to implement a deterministic calculation of a target function, with a success probability impossible to achieve using any multi-valued classical circuit that has a fixed depth, and thus converting the large quantum--classical ratio to a strong quantum advantage. 

\subsection{The original GHZ paradox in the probability form} 

Consider the following Pauli product measurements on a three-qubit system: $\mathcal{M}_1=\sigma_x^{(1)}\sigma_y^{(2)}\sigma_y^{(3)}$, $\mathcal{M}_2=\sigma_y^{(1)}\sigma_x^{(2)}\sigma_y^{(3)}$, $\mathcal{M}_3=\sigma_y^{(1)}\sigma_y^{(2)}\sigma_x^{(3)}$, and $\mathcal{M}_4=\sigma_x^{(1)}\sigma_x^{(2)}\sigma_x^{(3)}$. If the system is initiated at a state so the measurement outcome of $\mathcal{M}_1, \mathcal{M}_2$ and $\mathcal{M}_3$ are all $+1$, then a noncontextual theory will predict the measurement of $\mathcal{M}_4$ on the system definitely returns $+1$, because the assumption of noncontextuality means the outcome of measuring $\sigma_x^{(1)}$ will not depend on whether it is measured with $\sigma_x^{(2)}\sigma_x^{(3)}$ as in $\mathcal{M}_0$ or $\sigma_y^{(2)}\sigma_y^{(3)}$ as in $\mathcal{M}_1$; this is similar for $\sigma_x^{(2)},\sigma_x^{(3)}$. Taking into account the involutory of Pauli operators, the outcome of $\mathcal{M}_0$ in such theories will be the product of $\mathcal{M}_1, \mathcal{M}_2$ and $\mathcal{M}_3$ that is $+1$. However, quantum theory is contextual: the only three-qubit state $\ket{\psi}$ giving $\braket{{\cal M}_1}_\psi=\braket{{\cal M}_2}_\psi=\braket{{\cal M}_3}_\psi=+1$ is the GHZ state $\ket{\psi}=\ket{\rm GHZ}=(\ket{000}-\ket{111})/\sqrt{2}$, which satisfies $\braket{{\cal M}_4}_{\rm GHZ}=-1$.

Each of the four Pauli product measurements is composed of three local dichotomic measurements. If the outcomes of ${\cal M}_1$ through ${\cal M}_4$ are specified as in the above GHZ paradox, then the local dichotomic measurements corresponding to each Pauli product measurement will have four possible combinations. For example, ${\cal M}_4=-1$ implies either $\sigma_x^{(1)}=\sigma_x^{(2)}=\sigma_x^{(3)}=-1$, or that one and only one local measurement among $\sigma_x^{(1)},\sigma_x^{(2)}$ and $\sigma_x^{(3)}$ evaluates to $-1$. Let $\Pi_\mu^{\pm (\nu)}=(\openone \pm\sigma_\mu^\nu)/2$, then the four elementary events can be expressed as:
\begin{align}
\begin{aligned}
    1|[4,1]:&=1\,|\,\Pi_x^{-(1)}\otimes\Pi_x^{-(2)}\otimes\Pi_x^{-(3)}, \\
    1|[4,2]:&=1\,|\,\Pi_x^{-(1)}\otimes\Pi_x^{+(2)}\otimes\Pi_x^{+(3)}, \\
    1|[4,3]:&=1\,|\,\Pi_x^{+(1)}\otimes\Pi_x^{-(2)}\otimes\Pi_x^{+(3)}, \\
    1|[4,4]:&=1\,|\,\Pi_x^{+(1)}\otimes\Pi_x^{+(2)}\otimes\Pi_x^{-(3)}.
\end{aligned}
\end{align}
In the same vein, we can define the elementary events $1|[j,k], j\in\{1,2,3\}, k\in[1,4]$. The exclusivity graph for all these events, as discussed in the main text, is the graph complement of the Shrikhande graph, so the total number of events that are allowed to happen according to any noncontextual model is no more than 3 for any initial state. That is, given $\sum_{k=1}^4\Pr(1|[j,k])=1, j\in\{1,2,3\}$ it must be 
\begin{align}
    \sum_{k=1}^4\Pr(1|[4,k])\overset{\rm NCHV}{=}0.
\end{align} However, according to the calculated quantum expectations, the total probabilities of all the four groups of events will saturate the upper bound allowed by the principle of exclusivity, that is, 
\begin{align}
    \sum_{k=1}^4\Pr(1|[j,k])\overset{\rm Q}{=}1, j\in[1,4].
\end{align} This completes the transformation of the original GHZ paradox into the probability form as in the Eq.\,(1) in the main text.

\subsection{Quantum realization of the exclusivity graph}

Here we discuss how to obtain a detailed numeric setting for a set of rays that demonstrates the GHZ-type paradox from the exclusivity graph. Let $G$ be the exclusivity graph of interest with $|V(G)|=N$, where $V(G)$ is the vertex set of $G$. The Gram matrix $\mathbf{B}$ here is an $N\times N$ matrix associated with the orthogonal representation\,\cite{Lovasz89} of the graph complement of $G$. Specifically, it can be calculated using semidefinite programming from the dual form of the Lov\'asz optimization:
\begin{alignat}{3}
    {\max_\mathbf{B}}&\quad\tr(\mathbf{BJ}) \label{eq:lovasz-sdp}\\
    {\rm subject~to}&\quad\mathbf{B} \succcurlyeq 0 \quad\quad &&(\text{positive semidefiniteness}) \nonumber\\
    &\quad B_{ij} = 0,\ \forall\ (i,j)\in E(G) \quad\quad &&(\text{orthogonality}) \nonumber\\
    &\quad\tr(\mathbf{B}) = 1 \quad\quad &&(\text{normalisation}). \nonumber
\end{alignat}
Here, $\mathbf{J}$ is an $N\times N$ matrix with all of its entries being 1, $(\cdot)\succcurlyeq0$ means the matrix on the left-hand side of the operator is positive semidefinite, and $E(G)$ is the edge set of $G$. The second constraint is due to that when $(i,j)\in E(G)$, the rank-1 projectors $\Pi_i=\ket{v_i}\bra{v_i}$ and $\Pi_j=\ket{v_j}\bra{v_j}$ corresponding to the vertices $i$ and $j$ are exclusive: $\Pi_i\Pi_j=0$, so the rays $\ket{v_i}$ and $\ket{v_j}$ must be orthogonal, that is, $B_{ij}=\braket{v_j|v_i}=0$. The maximum of the objective function is precisely the Lov\'asz number of the exclusivity graph, $\vartheta(G)$. By running the Lov\'asz optimization and calculating $\vartheta(G)$, the Gram matrix is automatically obtained. In our study, we used the Python package \texttt{cvxopt} as the optimizer for the semidefinite programming. Since the optimizer is a black-box model we can only get access to the final output matrix $\mathbf{B}$. After several trials of optimization, we always get the same Gram matrix in the Extended Data Figure 1, and the solution has ${\rm Rank}(\mathbf{B})=37$. 

To work out the set of rays $\mathbf{r}$ that give rise to the Gram matrix, we will need to write $\mathbf{B}$ as the product of a real matrix and its transpose:
\begin{align}
    \mathbf{B} = \mathbf{r}\mathbf{r}^{\rm T}.
\end{align}
The objective is akin to the Cholesky decomposition where the result is limited to be a lower triangle matrix. However, the Cholesky decomposition only works for symmetric and positive-definite matrices and thus would not work directly for the gram matrix here which is not full-rank. Therefore, we take the first 37 rows and columns of $\mathbf{B}$ and denote it as $\tilde{\mathbf{B}}$. We can check that $\tilde{\mathbf{B}}$ is nondegenerate and thus the Cholesky decomposition can be run on it. This way, we obtain the first 37 rays. Further, Each of the other 20 rays can be parameterized as a 37-dimensional vector. As the Gram matrix specified their inner product with the existing rays, it is possible to write these relations as 20 sets of rank-37 linear equations. Due to that $\tilde{\mathbf{B}}$ is full-rank, all these linear systems are uniquely determined, and the last 20 rays can be obtained in a deterministic manner using Gaussian elimination.

For the Lov\'asz optimization of the complementary of the Perkel graph, the solution of $\mathbf{B}$ is robust probably because it has a symmetric form and preserves the symmetry of the graph. We cannot prove the optimality of the solution found in terms of dimensionality, that is, we cannot exclude the possibility that an orthogonal representation of the Perkel graph with a Gram matrix rank below 37 exists and saturates the Lov\'asz number. The primary difficulty is because a rank-constrained optimization is generally not convex and it is not possible to find a solution with semidefinite programming\,\cite{xdyu22}. We have run several trials of see-saw optimization and cannot find a rank-36 matrix $\mathbf{B}$ saturating the upper bound of the target function and satisfying all the conditions in \eqref{eq:lovasz-sdp}. 

The best estimation of the lowest dimensional realization seems to come from the existence of the graph's orthogonal realization. The lowest dimension $d^*({\bar G})$ to allow the orthogonal representation of a graph ${\bar G}$ to exist satisfies $\alpha({\bar G})\leqslant d^*({\bar G})\leqslant\chi({\bar G})$. In our case, setting ${\bar G}$ to be the Perkel graph, we have $\alpha({\bar G})=19, \chi({\bar G})=29$; therefore, some improvement for the dimension seems to be allowed, but the resulted orthogonal representation will still be of quite a high dimensionality. As such, it would not reduce the challenge for the experimental side by a lot. Instead, it is convenient to use the symmetric Gram matrix as found in the semidefinite programming and pursue an experimental observation of the high-dimensional GHZ-type contextuality.

\subsection{Pulse encoding}
\label{subsec:encoding}

In this subsection, we exemplify how a prepare-and-measure experiment is done in our experiment with a pulse-sequence encoding. We would like to evaluate the projection probability $|\braket{\revc|\boldsymbol{a}}|^2$ of the following ray:
\begin{align}
    \boldsymbol{a} = \bigg\{ \underbrace{\frac{1}{\sqrt{19}},\, \frac{1}{\sqrt{19}},\, \ldots,\, \frac{1}{\sqrt{19}}}_{\times 19},\ \underbrace{\phantom{\frac{1}{\sqrt{19}}} \hspace{-22pt} 0,\, 0,\, \ldots,\, 0}_{\times 18} \bigg\},
\end{align}
on the measurement basis:
\begin{align}
    \revc = \{-&0.110,0.170,0.091,0.257,0.040,0.019,-0.427,-0.131,-0.276,0.387, \nonumber\\ &0.112,-0.252,0.157,0.233,-0.010,-0.023,0.119,-0.147,-0.210,0,0, \nonumber\\ &0,0,0,0,0,0,0,0,0,0,0,0.424,-0.140,0.074,0.028,-0.123\}.
\end{align}
The basis here is not any of the 57 rays in the orthogonal representation of the Perkel graph, but it is orthogonal to the rays $\#20$ through $\#38$. Thus, this projection probability is required to build the probability distribution on an orthonormal basis. Also, the first seven terms of the intrinsic pulse amplitude ejection ratio from the fiber ring, normalized against the first pulse, are: 
\begin{align}
    \boldsymbol{c_0} = \{1.,0.783,0.606,0.465,0.356,0.272,0.208\}.
\end{align}
For convenience, we also denote $\revc_0 = \{0.208,0.272,0.356,0.465,0.606,0.783.1.\}$ as the entry-reversed kernel of convolution. In Table \ref{tab:sm-defs}, we summarize the meaning of various definitions used in this section.

\begin{table}[b!]
    \centering
    \begin{tabular}{c|l}
    \toprule
    Symbol & \hspace{.4\columnwidth} Definition \\
    \midrule
        $\boldsymbol{a}$ ($\revc$) & Desired input state (measurement basis) \\
        $\boldsymbol{c}_0$ & The kernel of convolution, determined only by the fiber ring \\
        $\revc_0$ & The kernel of convolution, but with the entry written in a reversed direction\\
        $\boldsymbol{a}_k$ ($\revc_k$) & The $k$-th part of $\boldsymbol{a}$ ($\revc$) in the direct-sum expansion: $\boldsymbol{a} = \bigoplus_{k=1}^b \boldsymbol{a}_k$, $\revc = \bigoplus_{k=1}^b \revc_k$, ${\rm Dim}(\boldsymbol{a}_k) = {\rm Dim}(\revc_k)$ \\
        $\revc_k^\downarrow$ & A list with the same elements as $\revc_k$ but sorted in the absolute-value descending order\\
        $\boldsymbol{a}_k^\downarrow$ & A list with the same elements as $\boldsymbol{a}_k$ but sorted according to that $(\boldsymbol{a}_k^\downarrow)_m = (\boldsymbol{a}_k)_n$ if $(\revc_k^\downarrow)_m = (\revc_k)_n$\\
        $V_k^\downarrow$ ($V_k^{\not\downarrow}$) & Modulation voltages for generating a pulse sequence with the amplitudes being $\braket{\revc_k^\downarrow|\boldsymbol{a}_k^\downarrow}$ ($\braket{\revc_k|\boldsymbol{a}_k}$) \\
        $\tilde{\boldsymbol{a}}$ & A ray reconstructed by measuring the amplitudes of all $\braket{\revc_k|\boldsymbol{a}_k}$'s: the theoretical value is $\tilde{\boldsymbol{a}}_k = \braket{\revc_k|\boldsymbol{a}_k}$ \\
        $\tilde{\revc}$ & An all-one vector: $\tilde{\revc} = \{1, 1, \ldots, 1\}$. Also, ${\rm Dim}(\tilde{\boldsymbol{a}}) = {\rm Dim}(\tilde{\revc}) = b$ \\
    \bottomrule
    \end{tabular}
    \caption{Summary of various symbol definitions used in the section\,\ref{subsec:encoding}.}
    \label{tab:sm-defs}
\end{table}

To measure the detection probability with the subspace method, the first step is to write the ray and basis into direct-sum decompositions:
\begin{align}
    \boldsymbol{a} = \bigoplus_{k=1}^6 \boldsymbol{a}_k, \quad \text{where} \ \ \boldsymbol{a}_1 = \boldsymbol{a}_2 &= \left\{\frac{1}{\sqrt{19}}, \frac{1}{\sqrt{19}}, \frac{1}{\sqrt{19}},\frac{1}{\sqrt{19}}, \frac{1}{\sqrt{19}}, \frac{1}{\sqrt{19}}, \frac{1}{\sqrt{19}}\right\}, \\
    \nonumber \boldsymbol{a}_3 &= \left\{\frac{1}{\sqrt{19}}, \frac{1}{\sqrt{19}}, \frac{1}{\sqrt{19}},\frac{1}{\sqrt{19}}, \frac{1}{\sqrt{19}}\right\}, \\ \nonumber  \boldsymbol{a}_4 &= \boldsymbol{a}_5 = \boldsymbol{a}_6 = \left\{0, 0, 0, 0, 0, 0\right\}.
\end{align}
and
\begin{align}
     \revc = \bigoplus_{k=1}^6 \revc_k, \quad \text{where} \ \ \revc_1 &= \{-0.110,0.170,0.091,0.257,0.040,0.019,-0.427\}, \\ \nonumber \revc_2 &= \{-0.131,-0.276,0.387,0.112,-0.252,0.157,0.233\}, \\ \nonumber \revc_3 &= \{-0.010,-0.023,0.119,-0.147,-0.210\}, \\ \nonumber \revc_4 &= \revc_5 = \{0, 0, 0, 0, 0, 0\}, \\ \nonumber \revc_6 &= \{0,0.424,-0.140,0.074,0.028,-0.123\}.
\end{align}
This way, the inner product can be expressed as $\braket{\revc|\boldsymbol{a}}=\textstyle\sum_{k=1}^b\braket{\revc_k|\boldsymbol{a}_k}$ as in the Eq.\,(7) in the main text.

\begin{figure}[b!]
    \centering
    \includegraphics[width = .45 \textwidth]{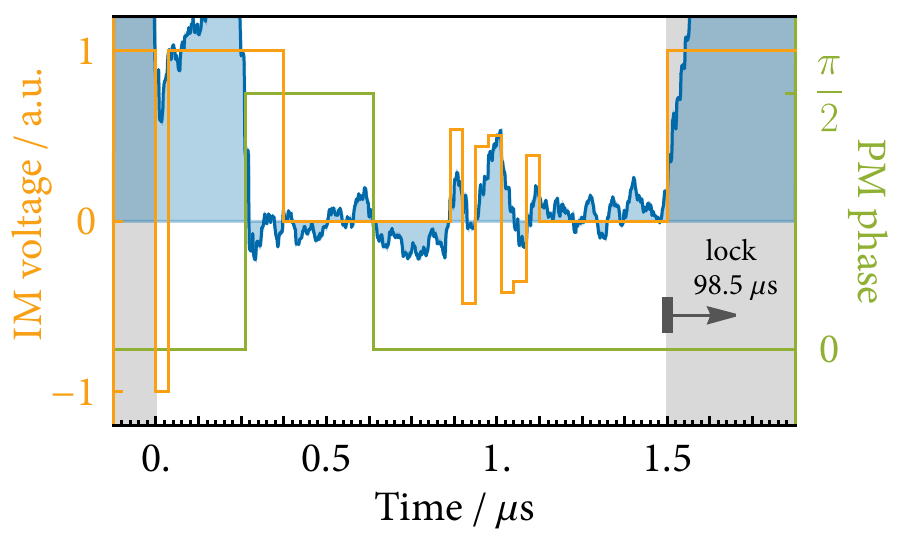}
    \hspace{.05\textwidth}
    \includegraphics[width = .45 \textwidth]{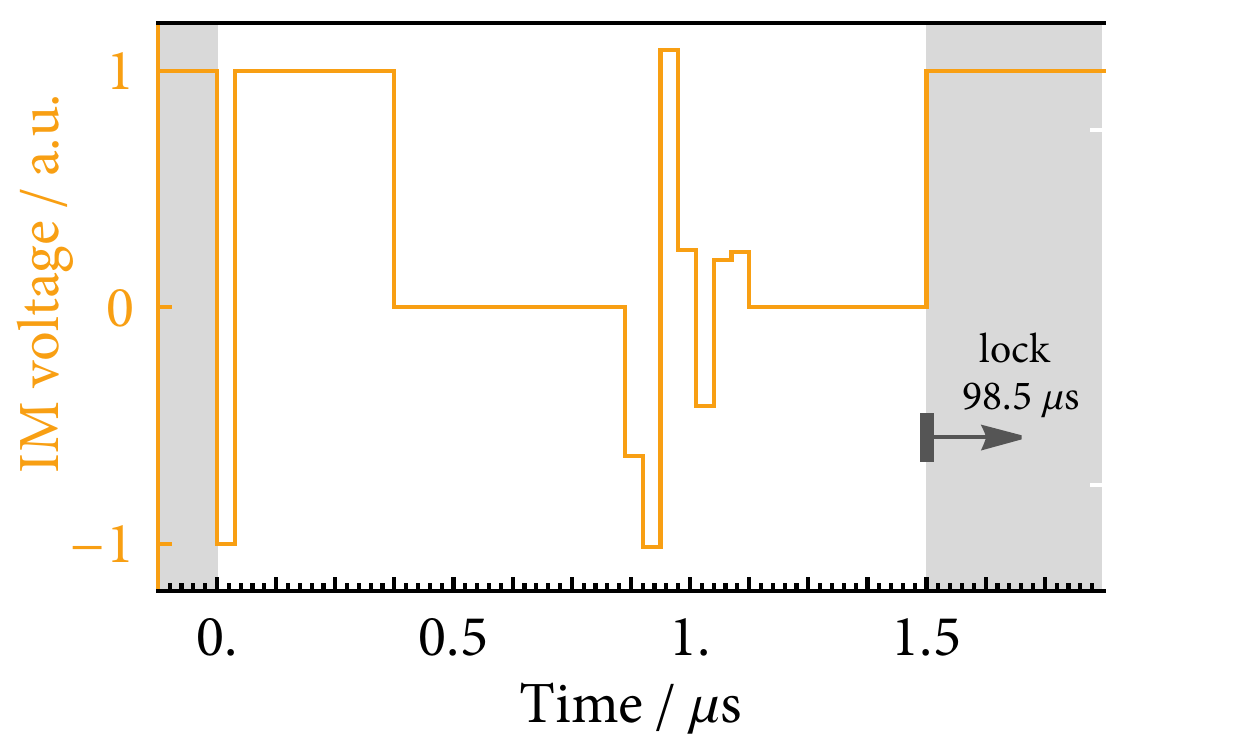}
    \hspace{-.05\textwidth}
    \caption{\textbf{Explanation of waveforms.} Left: The pulse encoding for evaluating the inner product $\braket{\revc_2|\boldsymbol{a}_2}$. Right: The effect of sorting the entries of $\revc\to\revc^\downarrow$. This is how the pulse for the intensity modulator needs to be if the entries are not sorted. Note the amplitude required to encode some of the entries exceeds the amplitude in the locking stage.}
    \label{figs:waveform}
\end{figure}

The second step is to construct the correct modulation signal for the intensity modulator to measure the inner products $\braket{\revc_k|\boldsymbol{a}_k}$. Throughout this part, we always make the intensity modulator work in a small transmission range, so the amplitude after the intensity modulator has a linear response to the modulation voltage. The modulation signal works by both creating the desired input state $\boldsymbol{a}$ and changing the natural convolution basis $\revc_0$ to the measurement basis $\revc_k, k\in[1, 6]$. To this purpose, the amplitude of the pulses in the $k$-th pulse sequence generated by the intensity modulator should be proportional to $V_k=\boldsymbol{a}_k\circ\revc_k\circ\revc_0^{\circ -1}$. Here, $\circ$ means Hadamard (entry-wise) product and $(\cdot)^{\circ -1}$ means Hadamard (entry-wise) inverse. In case the dimension of $\revc_0$ does not match that of $\boldsymbol{a}_k$ and $\revc_k$, the excess entries of $\revc_0$ are truncated. Additionally, to better utilize the linear range of the intensity modulator, we also sort the entries of the basis $\revc_k$ in an absolute-value descending order, hereafter denoted as $\revc_k^\downarrow$, and use $\revc_k^\downarrow$ to replace $\revc_k$ in the experiment. The entries of the input states are also adjusted accordingly. 

We use the second direct-sum subspace as an example. According to the definition above, the modulation voltage we applied on the intensity modulator, after sorting the entries, should be proportional to
\begin{align}
    V_2^\downarrow &= \boldsymbol{a}_2 \circ \revc_2^\downarrow \circ \revc_0^{\circ -1} \\
    &= \left\{\frac{1}{\sqrt{19}}, \frac{1}{\sqrt{19}}, \frac{1}{\sqrt{19}},\frac{1}{\sqrt{19}}, \frac{1}{\sqrt{19}}, \frac{1}{\sqrt{19}}, \frac{1}{\sqrt{19}}\right\} \circ \left\{\frac{0.112}{0.208}, -\frac{0.131}{0.272}, \frac{0.157}{0.356}, \frac{0.233}{0.465}, -\frac{0.252}{0.606}, -\frac{0.276}{0.783}, \frac{0.387}{1}\right\} \nonumber \\
    &= \{0.540,-0.481,0.441,0.501,-0.417,-0.352,0.387\}/\sqrt{19}. \nonumber
\end{align}
This corresponds to the waveform displayed in the Fig.\,3 in the main text. For convenience, we also show it here in the left subplot of Fig.\,\ref{figs:waveform}. On the other hand, if we did not implement the sorting procedure, the modulation voltage would become:
\begin{align}
    V_2^{\not \downarrow} &= \boldsymbol{a}_2 \circ \revc_2^{\not \downarrow} \circ \revc_0^{\circ -1} \\
    &= \left\{\frac{1}{\sqrt{19}}, \frac{1}{\sqrt{19}}, \frac{1}{\sqrt{19}},\frac{1}{\sqrt{19}}, \frac{1}{\sqrt{19}}, \frac{1}{\sqrt{19}}, \frac{1}{\sqrt{19}}\right\} \circ \left\{-\frac{0.131}{0.208}, -\frac{0.276}{0.272}, \frac{0.387}{0.356}, \frac{0.112}{0.465}, -\frac{0.252}{0.606}, \frac{0.157}{0.783}, \frac{0.233}{1}\right\} \nonumber \\
    &= \{-0.629,-1.01,1.09,0.241,-0.417,0.201,0.233\}/\sqrt{19}. \nonumber
\end{align}
This waveform, corresponding to the right subplot of Fig.\,\ref{figs:waveform}, would thus require larger modulation amplitudes for some of the entries, and become more susceptible to the non-perfect linear response of the modulator.

The final step is we register the homodyne amplitude when the pulses from all the seven time-bins have entered the ring, and use this for all the subspaces to create a new pulsed state. In our case, we measured these six homodyne amplitudes to be: $\tilde{\boldsymbol{a}} = \{1.389, 4.389, -1.650, 0.756, 0.756, 5.66316\}$. The values are direct readouts from the oscilloscope trace and no normalization is needed. By taking this as a new ray and convolving it with a unit kernel $\tilde{\revc}=\{1, 1, 1, 1, 1, 1\}$ using again the above-mentioned procedure, we obtained the final unnormalized inner product $\braket{\tilde{\revc}|\tilde{\boldsymbol{a}}}$. Normalizing the square of the inner product across an orthonormal basis yields the detection probabilities of different measurements.

\subsection{Proposed improvement of the setup}

\begin{figure}[h!]
    \centering
    \includegraphics[width=0.98\textwidth]{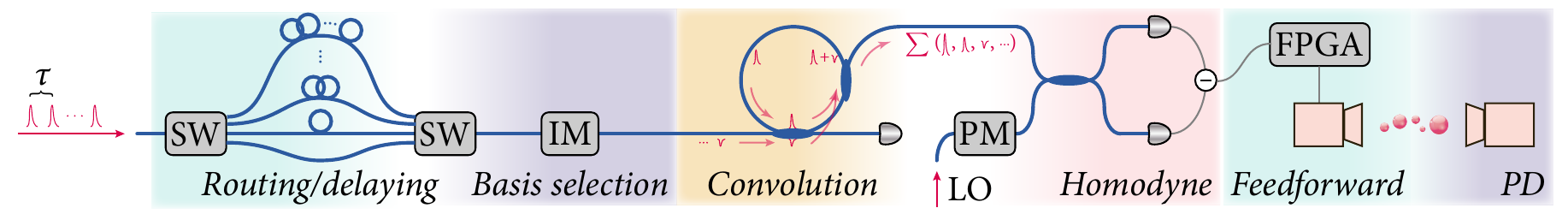}
    \caption{Proposed setup modification for combining different measurement stages, ``discretizing'' the measurement outcome, producing Wigner negativity and making the experiment meet the prerequisites of the NCHV theory. Comparing with the main setup, the modification is twofold: a delay and switch stage is added, and the last stage of measurement uses feedforward and photodetection. SW switch, PD photodetector.}
    \label{fig:s-setup-mod}
\end{figure}

As we mentioned in the main text, the current setup is unable to observe contextuality at the event-level by producing individual events and measure their probability. It is thus not fully compatible with the requirements of a noncontextual hidden-variable theory, However, we believe that it is possible to modify the setup to accommodate the requirements of the standard noncontextual model and, at the same time, manifest contextuality. The key point is to introduce an operation in the measurement procedure that (1) produces event-level outcomes and (2) induces negativity to the Wigner function. A photodetection process at the last stage of the measurement (labeled as ``Round b+1'' in Fig. 2c in the main text) will satisfy both of the requirements and promote the current experiment to a bona fide contextuality test. This extended schematic setup is shown in Fig.\,\ref{fig:s-setup-mod}.

The setup in Fig.\,\ref{fig:s-setup-mod} contains three more stages compared to our setup. The first is that the encoding of the high-dimensional state should happen on the different segments of a single pulse train. At the beginning of the setup, a routing and delaying unit is added to enable the measurement from round 1 to round $b$ in the main setup to happen one by one in the same setup. The switch and routing between MHz-frequency pulses are possible with commercial electro-optical modulators\,\cite{Xanadu22} and the fiber delay lines for separating different segments of pulses will be of moderate length. Moreover, it appears that the relative phases between different delay lines does not need to be locked: by adding a pilot wave in front of every segment, the phase between them can be individually determined, which will greatly reduce the challenge due to the phase locking of long fibers. 

The second point of change is that, after the homodyne detection, the outcomes would be sent to a field programmable gate array (FPGA) for real-time processing, thus the $b$-rounds of homodyne measurement outcomes would be stored inside the FPGA, and the FPGA would make another round of preparation process, but this time with the quantum state encoded on a single photon. A photodetection process at the end of the setup would then project the input state on a single basis. This way, the measurement stage starting from the ``basis'' selection until the photodetection would constitute a mapping from a Hilbert space to an event-level outcome, and conform to the requirements of a NCHV theory. Further, changing the power detection to photodetection will maintain the probability, so we would expect to observe the same amount of contextuality with the modified setup. We consider encoding the photonic qudit on the spatial mode degree of freedom using spatial light modulation as a possible approach\,\cite{zhliu23phcp}. This kind of setup is also compatible with a FPGA-based control system\,\cite{Canas14,Lima18}. Finally, the FPGA for doing the necessary calculation and implementing the feedforward operation itself was available in the setting of coherent Ising machine\,\cite{McMahon16, Inagaki16}.

\subsection{Full proof of the Theorem 3}

In this subsection, we prove that there is no strongly regular graph with clique number 2 and chromatic number 3 such that the Lov\'{a}sz number of its complement graph is 3. By the Theorem 1 in the main text, it means no strongly regular graph hosts a three-context GHZ-type paradox.

A strongly regular graph with four parameters $n,k,a$ and $c,$ denoted by ${\rm SRG}(n,k,a,c),$ is a $n$-vertex $k$-regular graph such that every two adjacent vertices have $a$ common neighbors, and that every two non-adjacent vertices have $c$ common neighbors. The theory of strongly regular graphs was introduced by Bose\,\cite{Bose63} in 1963 and was then used in many different mathematical fields, such as partial geometry\,\cite{Bose63}, group theory\,\cite{Higman64} and coding theory\,\cite{Delsarte72}.

Before proving the main lemma, let us list out some necessary symbols. For any graph $G$, we use $A(G)$, $\alpha(G)$, $\omega(G)$, $\chi(G)$ and $\vartheta(G)$ to denote its adjacent matrix, independence number, clique number, chromatic number and Lov\'{a}sz number, respectively. Suppose that there exists a strongly regular graph $G={\rm SRG}(n,k,a,c)$ satisfying $\omega(G)=2, \chi(G)=3$ and $\vartheta(\bar G)=3$. Note that as $\omega(G)=2,$ we have $a=0.$ 

We will first establish an inequality that $c\leqslant k\leqslant 2c+3.$	
Let $A=A(G)$. We will consider the entries of $A^2.$ For any two vertices $i$ and $j,$ $(A^{2})_{ij}$ is the number of walks of length 2 which starts at $i$ and ends at $j.$ So for adjacent vertices $i$ and $j,$ $(A^{2})_{ij}=0$ because $i$ and $j$ has no common neighbors. For non-adjacent vertices $i$ and $j,$ $(A^{2})_{ij}=c$ because $i$ and $j$ has exactly $c$ common neighbors. And for any vertex $i,$ $(A^{2})_{ii}=k$ because $G$ is a $k$-regular graph. Therefore, we have $A^{2}=c(J-A)+(k-c)I,$ where $J$ is the $n\times n$ matrix with all entries equal to 1 and $I$ is the $n\times n$ identity matrix. This implies that $A$ has exactly three different eigenvalues $k,\lambda_{1},\lambda_{2}$ such that $\lambda_{1}$ and $\lambda_{2}$ are the roots of $\lambda^{2}+c\lambda-(k-c)=0.$ Therefore
\begin{align}
    \lambda_{1}=\frac{-c-\sqrt{c^{2}+4(k-c)}}{2}, \text{\quad and\quad} \lambda_{2}=\frac{-c+\sqrt{c^{2}+4(k-c)}}{2}.
\end{align} 
The famous Hoffman--Delsarte\,\cite{Cvetkovic80,Delsarte73} eigenvalue bound says that for $n$-vertex $k$-regular graph $H,$ we have $\alpha(H)\leqslant\frac{-\tau}{k-\tau}n,$ where $\tau$ is the smallest eigenvalue of $A(H)$. Using this bound, we have 
\begin{align}
    \alpha(G)\leqslant\frac{-\lambda_{1}}{k-\lambda_{1}}n.
\end{align}
Also, since $\alpha(G)\chi(G)\geqslant n,$ we get another inequality 
\begin{align}
    3=\chi(G)\geqslant\frac{k-\lambda_{1}}{-\lambda_{1}}=\frac{2k+c+\sqrt{c^{2}+4(k-c)}}{c+\sqrt{c^{2}+4(k-c)}}.    
\end{align}
By solving this inequality, we have $c+2-\sqrt{c^2+4}\leqslant k\leqslant c+2+\sqrt{c^{2}+4}.$ Since $k,c$ are non-negative integers and $c\leqslant k,$ we now have 
\begin{align}
    c\leqslant k\leqslant 2c+3.
\end{align}

Next we show that in fact $k$ only can be two possible integer values, namely $k\in \{c,2c+1\}$, by considering the multiplicities of $\lambda_{1}$ and $\lambda_{2}.$ 
Let $m_{i}$ be the multiplicity of $\lambda_{i}$ for $i\in \{1,2\}.$ Since the multiplicity of $k$ is one, we have $m_{1}+m_{2}=n-1$; 
also by considering the trace of $A$, we get $m_{1}\lambda_{1}+m_{2}\lambda_{2}+k=0$.  
Solving the above two equations, we obtain that 
\begin{align}
\begin{aligned}
    m_{1}=\frac{(n-1)\lambda_{2}+k}{\lambda_{2}-\lambda_{1}}=\frac{n-1}{2}+\frac{-k^{2}+(2-c)k}{2\sqrt{c^{2}+4(k-c)}}.
\end{aligned}
\end{align}

As $m_{1}$ is an integer, we conclude that $\sqrt{c^{2}+4(k-c)}$ must be an integer.
Using $c\leqslant k\leqslant 2c+3,$ we have 
\begin{align}
    c\leqslant \sqrt{c^{2}+4(k-c)} \leqslant\sqrt{c^{2}+4c+12}<c+4.
\end{align}
So $\sqrt{c^{2}+4(k-c)}\in \{c, c+1, c+2, c+3\}$. An easy analysis shows that $k\in\{c,2c+1\}.$

By considering the number of paths of length 2 with a fixed an endpoint in $G$, we have an equality $(n-k-1)c=k(k-1).$ Therefore, $n=2c$ when $k=c,$ and $n=6c+4$ when $k=2c+1.$
That is, $G$ is either SRG$(2c,c,0,c)$ or SRG$(6c+4,2c+1,0,c)$.
In the former case, $G=$SRG$(2c,c,0,c)$ must be a complete bipartite graph such that each part has exactly $c$ vertices. Therefore $\chi(G)=2,$ a contradiction. 
So from now on, we only need to consider the later case $G$=SRG$(6c+4,2c+1,0,c)$.

We first consider the cases when $c\leqslant 2$.	
If $c=0,$ then $G$ is exactly a single edge, implying that $\chi(G)=2,$ a contradiction. 
If $c=1,$ then $G=$SRG$(10,3,0,1)$ is unique, i.e., the famous Peterson graph\,\cite{Petersen}. 
So we have $\vartheta(\bar G)=\frac{5}{2}$\,\cite{Lovasz79} which is again a contradiction. 
If $c=2,$ then $G=$SRG$(16,5,0,2)$ is also unique, i.e., the famous Clebsch graph\,\cite{Clebsch}. In this case, we have $\chi(G)=4,$ a contradiction.
	
It remains to show that $G$=SRG$(6c+4,2c+1,0,c)$ does not exist for $c\geqslant 3$.
Suppose such $G$ exists. Take an arbitrary vertex $v$ in $G$. Let $N(v)$ denote the set of neighbors of $v,$ and $V_{1}$ denote the non-neighbors of $v.$ 
As $G$ has no triangles, $N(v)$ is an independent set of size $2c+1$ and $|V_{1}|=4c+2.$ 
By the properties of strongly regularity of $G$, any vertex $w$ in $N(v)$ has exactly $2c$ neighbors in $V_{1}$, 
and any vertex $u$ in $V_{1}$ has exactly $c$ neighbors in $N(v)$ and $c+1$ neighbors in $V_{1}$.
	
Fix a vertex $u$ in $V_{1}.$ Because $c\geqslant 3,$ we can take three different neighbors $w_{1},w_{2},w_{3}$ of $u$ in $N(v).$ 
Let $W_{i}$ denote the set of the neighbors of $w_{i}$ except $v$ for all $i\in\{1,2,3\}.$ 
Then $W_{i}\subseteq V_1$ and $W_i$ is an independent set of size $2c$ for all $i\in\{1,2,3\}.$ 
Moreover, if $z$ is a neighbor of $u$ in $V_{1},$ $z$ cannot be a neighbor of either $w_{1},w_{2}$ or $w_{3}.$ 
So we have $N(u)\cap V_{1}\subseteq V_{1}\setminus(W_{1}\cup W_{2}\cup W_{3}).$ 
Since any two non-adjacent vertices in $G$ have exactly $c$ common neighbors and $v$ is a common neighbor of $w_{1},w_{2}$ and $w_{3},$ 
we have $|W_{1}\cap W_{2}|=c-1$ and $|W_{3}\cap(W_{1}\cup W_{2})|\leqslant |W_{3}\cap W_{1}|+|W_{3}\cap W_{2}|=2c-2.$ 
Therefore $|W_{1}\cup W_{2}|=3c+1$ and thus $|W_{1}\cup W_{2}\cup W_{3}|\geqslant (3c+1)+2c-(2c-2)=3c+3.$
Finally, we can reach a contradiction by the following inequality:
\begin{align}
\begin{aligned}
    c+1&=|N(u)\cap V_{1}|\leqslant |V_{1}\setminus(W_{1}\cup W_{2}\cup W_{3})| \\
    &\leqslant (4c+2)-(3c+3)=c-1. 
\end{aligned}
\end{align}
The proof is now complete. 


\end{document}